\documentclass[conference]{IEEEtran}

\usepackage{amsmath}
\renewcommand{\baselinestretch}{0.85}

\usepackage{mathrsfs}
\usepackage{float}
\usepackage{courier}
\usepackage{cite}
\usepackage{url}
\usepackage{graphicx}
\usepackage{listings}
\usepackage{subcaption}
\usepackage{lipsum}
\usepackage[english]{babel}

\newcommand{\Tabref}[1]{Table~\ref{#1}}
\newcommand{\Figref}[1]{Figure~\ref{#1}}

\newcommand{\Sectref}[1]{Section~\ref{#1}}

\newcommand{\comment}[1]{}
\renewcommand\IEEEkeywordsname{Keywords}

\newtheorem{remark}{Remark}

\begin{document}
\title{ \LARGE \bf A Wireless Intrusion Detection System for 802.11 WPA3 Networks}
\author {Neil Dalal, Nadeem Akhtar, Anubhav Gupta, Nikhil Karamchandani, Gaurav S. Kasbekar, and Jatin Parekh}

\maketitle{}
{\renewcommand{\thefootnote}{} \footnotetext{N. Dalal, N. Karamchandani and G. Kasbekar are with the Department of Electrical Engineering, Indian Institute of Technology (IIT) Bombay, Mumbai, India. N. Akhtar, A. Gupta and J. Parekh are with Arista Networks. Inc. Their email addresses are neildalal125@gmail.com, nikhilk@ee.iitb.ac.in, gskasbekar@ee.iitb.ac.in, nadeem.akhtar@arista.com, 	
anubhav.gupta@arista.com, and jatin.parekh@arista.com, respectively. Part of this work was sponsored by Arista Networks. Inc}}

\begin{abstract}
Wi-Fi (802.11) networks have become an essential part of our daily lives; hence, their security is of utmost importance. However, Wi-Fi Protected Access 3 (WPA3), the latest security certification for 802.11 standards, has recently been shown to be vulnerable to several attacks. In this paper, we first describe the attacks on WPA3 networks that have been reported in prior work; additionally, we show that a deauthentication attack and a beacon flood attack, known to be possible on a WPA2 network, are still possible with WPA3. We launch and test all the above (a total of nine) attacks using a testbed that contains an enterprise  Access Point (AP) and Intrusion Detection System (IDS). Our experimental results show that the AP is vulnerable to eight out of the nine attacks and the IDS is unable to detect any of them. We propose a design for a signature-based IDS, which incorporates techniques to detect all the above attacks. Also, we implement these techniques on our testbed and verify that our IDS is able to successfully detect all the above attacks. We provide schemes for mitigating the impact of the above attacks once they are detected. We make the code to perform the above attacks as well as that of our IDS publicly available, so that it can be used for future work
by the research community at large. 
\end{abstract}


\section{Introduction}\label{Intro}
Today, Wi-Fi has become an essential part of our daily lives. More than one billion Wi-Fi Access Points (APs) exist, which provide internet services to over a hundred billion consumer devices such as smartphones, tablets, laptops, desktops, IoT devices, etc.~\cite{Pahlavan2021}. With such a large and growing number of  users and devices, the security of Wi-Fi networks  has become of utmost importance.

The Wi-Fi 802.11 standard has seen several revisions over the years since it was first defined in 1997~\cite{1997ieee}. But most of these revisions have concentrated on performance improvements, and only a few were designed to enhance security. While the 802.11 protocols are standards defined by IEEE, the Wi-Fi Alliance issues security certifications based on IEEE's work. Wired Equivalent Privacy (WEP), Wi-Fi Protected Access (WPA), and WPA2 are its previous security certifications, and now WPA3 is the latest one released in 2018~\cite{wpa3}. From June 2020 onwards, it has been mandatory for all new devices to support WPA3.

WPA2, which until recently was considered reasonably secure, has been shown to have a number of severe vulnerabilities and protocol loopholes~\cite{krack,dos,empirical}. Although WPA3 fixes these shortcomings of WPA2, some of the design and implementation flaws of WPA3 that allow denial-of-service (DoS), side-channel and downgrade attacks have been revealed in~\cite{dragon,depriv,badtoken}. Even after carefully designing standards, there might be vulnerabilities present, which an attacker can exploit. Hence, there is a need for an \emph{Intrusion Detection System} (IDS) to monitor the network as a second line of defense and raise alerts in real-time when an attacker is exploiting such vulnerabilities~\cite{ids_need}. Once an attack alert is raised, a security expert can inspect it more closely and execute steps to locate and block the attacker. There has been plenty of research on building an IDS for 802.11 networks~\cite{empirical,awid,awid2,semi,nsl1,nsl-kdd,gprs1,ids_need}, but none focuses specifically on WPA3. In this paper, we describe the design and implementation of an IDS that has been specifically designed for detecting attacks on WPA3 networks.

For performing various attacks on a wireless network, Aircrack-ng~\cite{aircrack}, MDK3~\cite{mdk3} and Metasploit~\cite{metasploit} are popular publicly available tools that network researchers have long used. However, none of these provide support for performing attacks on WPA3 networks. A more recent work~\cite{advcomm} demonstrates how advanced attacks on Wi-Fi networks can be carried out using cheap commodity hardware. In~\Sectref{attack}, we explain how we execute the known attacks on WPA3 networks, in order to test the efficacy of our IDS. 

Our contributions are as follows: \\
1) We describe the attacks on WPA3 networks that have been reported in prior work. Additionally, we show that a deauthentication attack and a beacon flood attack (Sections~\ref{deauth} and~\ref{beaconfld}), known to be possible on a WPA2 network, are still possible with WPA3. \\
2) We launch and test all the above (a total of nine) attacks using a testbed that contains an enterprise-grade AP supporting WPA3 and IDS that monitors our network for any malicious behavior. Our experimental results show that  the
enterprise AP we use is vulnerable to eight out of the nine attacks and the enterprise IDS is unable to detect any of these nine attacks. We also provide the steps and code to perform each of the above attacks. The files are available in our GitHub repository~\cite{gitlink}, which can be used for future work by the research community at large. \\
3) We propose a design for a \emph{signature-based IDS}~\cite{axel}, which incorporates techniques to detect all the above attacks. Also, we implement these techniques on our testbed and verify that \emph{our IDS is able to successfully detect all the above attacks}. The code of our IDS is also available at~\cite{gitlink}. \\
4) Finally, we provide schemes for mitigating the impact of the above attacks once they are detected.

The rest of the paper is organized as follows.~\Sectref{ReWo} provides a review of related prior work.~\Sectref{Network} describes the network model and  objectives of this paper.~\Sectref{attack} describes and explains how to perform various attacks on WPA3 networks.~\Sectref{sigdet} provides techniques to build an IDS to detect all of these attacks.~\Sectref{next} provides some schemes for mitigation of the impact of the above attacks. We present our experimental results in~\Sectref{exp}, and provide conclusions and directions for future research in~\Sectref{Con}.

\section{Related Work}\label{ReWo}
IDSs have been used for a long time for detecting attacks on wired as well as wireless networks; see~\cite{wired2} for a survey. Depending on the type of detection technique used, an IDS can broadly be classified as either anomaly or signature (also called misuse) based~\cite{ids1,ids2,history}.

There has been a lot of research on building a reliable high-performance anomaly-based IDS. In~\cite{nsl1}, the authors used deep learning methods for intrusion detection on the NSL-KDD datset~\cite{nsl-kdd}. This dataset is an improved version of the KDD Cup'99 dataset~\cite{kdd}, which was compiled in 1998 by DARPA by launching various attacks on a simulated wired network. The authors in~\cite{gprs1} used the GPRS database to train and test their IDS on. The GPRS database was built using an actual wireless network based on WEP. In~\cite{awid,awid2,semi}, the more recent AWID2 dataset~\cite{empirical} was used to build an anomaly-based IDS using different machine learning techniques. This AWID2 dataset released in 2016 is also based on WEP, but is much more comprehensive than the GPRS dataset.
Such machine learning based anomaly detection methods can detect previously unseen attacks, but they require a dataset (often labelled) to train their model on and have to deal with the problem of feature selection. Additionally, all the above mentioned datasets are outdated. They do not contain the attacks specific to WPA3, and most of the attacks they include are no longer a threat to WPA3 with its improved security. Even the latest AWID3 dataset~\cite{awid_21} released in 2021 is based on WPA2. In this paper and our GitHub repository~\cite{gitlink}, we provide procedures and codes to perform the known attacks on WPA3. We believe that \emph{these will aid research in building a new and updated dataset based on WPA3}, which in turn will help in developing improved IDSs.

While signature-based IDSs are not able to detect novel, zero-day attacks, they have the advantage that they provide a higher detection accuracy and lower false positive rate than anomaly-based ones~\cite{ids_tax}. Snort-Wireless~\cite{snort},  AirMagnet~\cite{airmag} and AirDefence~\cite{airdef} are some commercial signature-based IDSs for wireless networks. While these could be used to detect and prevent certain attacks in a WEP/ WPA/ WPA2 network, they are not updated to work with WPA3 networks. In~\Sectref{sigdet}, we present a design for a signature-based IDS capable of detecting attacks on WPA3. 

\section{Network Model and Objectives}\label{Network}

We set up an 802.11 wireless network running in infrastructure mode. All the APs and clients in the network support the WPA3 security standard. An AP is configured to run in either WPA3-Personal or WPA3-Personal Transition mode \cite{wpa3} as required for the analysis. We also set up an attack node capable of launching various attacks on this network. These attacks are known attacks specific to WPA3 (see Section~\ref{attack} for details).  Lastly, we also set up an IDS at the link layer that monitors our network for any malicious behavior. We check whether this IDS can successfully detect any of the attacks launched on the network. 

Our objectives are to perform the attacks, capture the packets exchanged during each attack, verify if they match the theoretically expected packet trace, and then identify the various attack signatures. Then, we seek to design a signature-based IDS capable of detecting all of these attacks. Next, our goal is to implement this design in code and test its performance on our 802.11 network setup.

\section{Changes in WPA3 from WPA2 and Known Attacks on WPA3 Networks}\label{attack}
First, in Section~\ref{SSC:changes:WPA3:WPA2}, we present the main changes introduced in WPA3, relative to WPA2, for enhancing the security. Then in Section~\ref{list_attack}, we describe all the currently known attacks on WPA3 and explain how to perform them.

\subsection{Changes in WPA3 from WPA2}
\label{SSC:changes:WPA3:WPA2}
\begin{figure}[h]
    \centering
    \includegraphics[scale = 0.24]{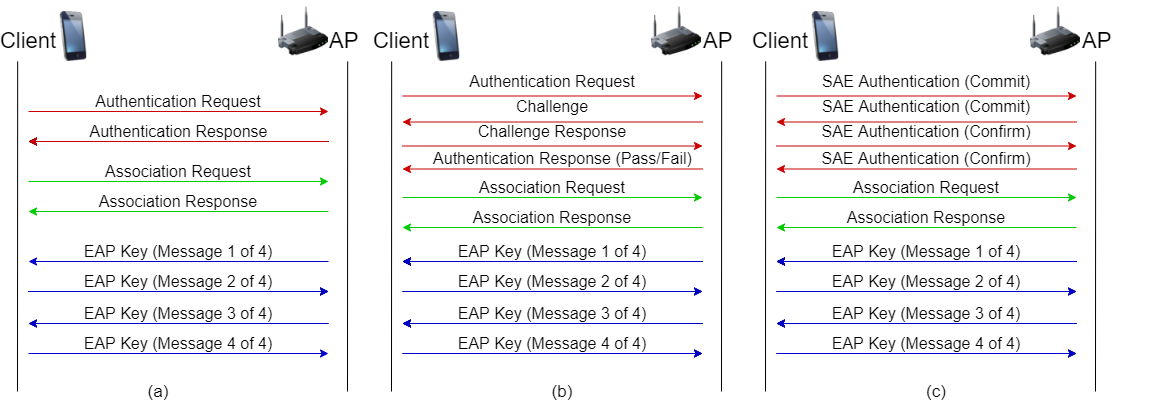}
    \caption{(a) Open System Authentication, (b) Shared Key Authentication, (c) Simultaneous Authentication of Equals (SAE)}
    \label{wpa3wpa2}
\end{figure}

\begin{figure}[h]
    \centering
    \includegraphics[scale = 0.45]{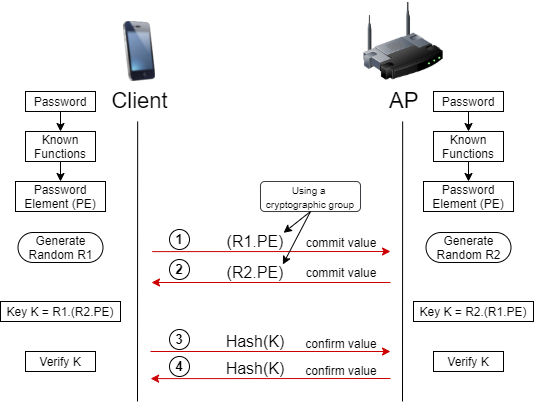}
    \caption{Simplified overview of SAE Authentication. Refer to~\cite{802112016} for detailed SAE mechanism.}
    \label{SAE_auth}
\end{figure}

For authentication between the client and the AP, previous security standards (WEP, WPA, WPA2) allowed either Open System authentication or Shared Key authentication~\cite{depriv}. On the other hand, WPA3 makes it compulsory to use Simultaneous Authentication of Equals (SAE) only~\cite{802112016}. 
\Figref{wpa3wpa2} (a) shows Open System Authentication (also called null authentication), wherein the parties simply exchange a request and a response message without any verification of the identity of the other party. 
\Figref{wpa3wpa2} (b) shows the use of Shared Key Authentication, wherein four messages are exchanged instead of two. The AP verifies the client by asking it to respond to a challenge;  WEP or RC4 is used to encrypt the challenge response. But WEP and RC4 have been proven to be breakable~\cite{wep1,wep2,wep3,wep4}, due to which Shared Key Authentication is considered a security risk.
\Figref{wpa3wpa2} (c) shows the SAE mechanism  (also referred to as Dragonfly handshake~\cite{dragon}), wherein again four messages are exchanged. The first two are called the commit frames and the subsequent two the confirm frames.
SAE achieves mutual authentication of client and AP and establishes secure keys to protect all information subsequently exchanged. \Figref{SAE_auth} gives a simplified overview of the SAE mechanism. As shown in the figure, in the two commit messages (no. 1 and 2), both the parties exchange a randomly generated secret element called the commit value used to generate a key. This exchange takes place using a cryptographic group and is similar to the Diffie-Hellman key exchange. Subsequently, in the  two confirm messages (no. 3 and 4), both parties exchange the hash of the keys that they have generated and verify (authenticate) that the other party is indeed legitimate.

Additionally, WPA3 makes it mandatory to use management frame protection (MFP) introduced in the IEEE 802.11w standard~\cite{80211w}. It provides data integrity and replay protection for some of the management frames. 
Making SAE and MFP mandatory prevents most of the attacks that were previously possible on WPA2 networks. The deauthentication flood attack \cite{deauthattack} is no longer possible as MFP protects the deauthentication frames and they cannot be spoofed. Fake authentication, offline dictionary attack, and other severe attacks \cite{krack,wpa2attack,vink} exploiting the vulnerabilities of Open System Authentication are also no longer possible as SAE is mandatory in WPA3. Still, there exist some known vulnerabilities in the WPA3 protocol, which an attacker can exploit~\cite{dragon},~\cite{depriv}. These attacks are described in Section~\ref{list_attack}. 

As shown in \Figref{wpa3wpa2}, after authentication and association, a 4-way handshake takes place using Extensible Authentication Protocol Over LAN (EAPOL)-Key frames~\cite{depriv}. This handshake is used to derive a Pairwise Transient Key (PTK)~\cite{depriv}. Particularly in the third message of the handshake (EAPOL Key Message 3), using the information from the Robust Security Network Element (RSNE)~\cite{depriv} the parties verify if both of them have the same security parameters of the network. RSNE contains information about the security protocols and parameters supported by the AP and is also shared within its beacons. If there is any mismatch between the RSNE held by the two parties, then the handshake is aborted. 

\subsection{Known Attacks on WPA3 Networks}\label{list_attack}
\Tabref{table1} lists the currently known attacks on WPA3. It comprises a total of nine attacks, of which the first seven are specific to WPA3 only. The last two attacks existed in WPA2 networks and we show that it is still possible to execute them on a WPA3 network. The table also includes two columns showing the results of our testing. 
It can be seen that the APs we used for this experiment are vulnerable to eight out of the nine attacks, and the IDS system was unable to detect any of these nine attacks. 
Next, we describe the attacks listed in Table~\ref{table1} and explain how to perform them.

\begin{table*}
\begin{center}
\caption{\label{tab:1}List of known attacks on WPA3 networks}
\begin{tabular}{ |c |c | c| c| c| c| } 
 \hline
 No. & Attack & Impact of attack & AP vulnerable? & IDS Result & Source\\
 \hline
 \hline
 1) & SAE Authentication flood attack & Denial-of-service & Yes & Not Detected & ~\cite{dragon}\\ 
 \hline
 2) & Try to make a client use WPA2 & Denial-of-service  & Yes & Not Detected & ~\cite{depriv}\\
 \hline
 3) & Downgrade to WPA2 attack & Client connects using WPA2 instead of WPA3 & Yes & Not Detected & ~\cite{dragon}\\
 \hline
 4) & SAE commit out of range attack & Denial-of-service  & Yes & Not Detected & ~\cite{depriv}\\
 \hline
 5) & SAE unsupported group attack & Denial-of-service  & Yes & Not Detected & ~\cite{depriv}\\ 
 \hline
 6) & Downgrade group attack & Less secure connection & Yes & Not Detected & ~\cite{dragon}\\
 \hline
 7) & Timing side channel attack & Leaks information about the password & No & Not Detected & ~\cite{dragon}\\ 
 \hline
 8) & Deauthentiction attack & Denial-of-service  & Yes & Not Detected & This Paper\\ 
 \hline
 9) & Beacon/ Probe frames flood attack & Confuse clients in trying to find the legitimate AP & Yes & Not Detected & This paper\\
 \hline
\end{tabular}
\label{table1}
\end{center}
\end{table*}

\subsubsection{SAE Authentication Flood Attack}\label{auth}\hfill\\
\textbf{Source:} This attack is from~\cite{dragon}.\\
\textbf{Impact:} Causes Denial-of-service (DoS)-- prevents clients from connecting to the network. \\
\textbf{Our test result:} AP vulnerable and IDS does not detect it.\\
\textbf{Short Description:} For authentication, WPA3 mandates the use of the SAE mechanism. In the operation of SAE, as shown in \Figref{SAE_auth}, to generate commit message no. 2, an AP needs to first compute a `password element' (PE). Calculating this PE is computationally expensive and this fact can be exploited by an attacker: By flooding the AP with SAE connection requests, the AP resources can get exhausted trying to calculate the PE and responding to each of the requests. This prevents the AP from serving newer clients, thus causing DoS. It also wastes airtime, thus reducing available bandwidth.
This attack is explained in detail in~\cite{dragon}.\\
\textbf{How to launch:} The authors of~\cite{dragon} provide a tool to execute this attack and also explain in detail how to use it in~\cite{dragondrain}.

\comment{
\begin{figure}[t!]
  \centering  
  \begin{minipage}[b]{0.5\linewidth}
        \centering

        \resizebox{1.0\columnwidth}{!}{\includegraphics{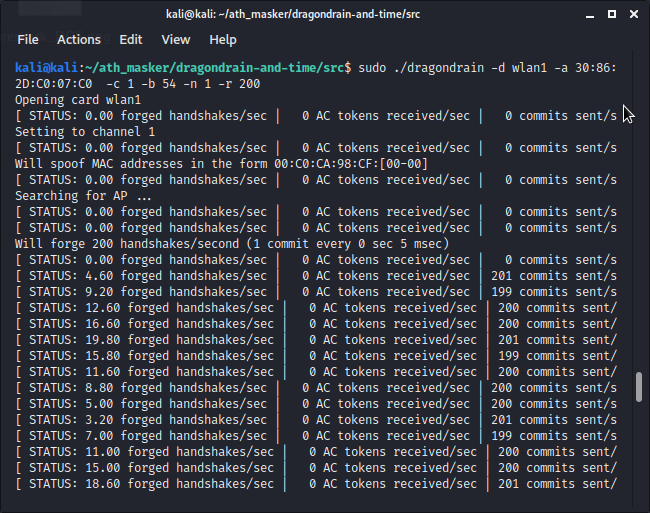}}
         %\caption{$W_{e}=W_{p}=W_{n}=1$}
%    	\label{Avg_Tp_Same_Wt} 
 \end{minipage}%
 ~
\begin{minipage}[b]{0.5\linewidth}
  \centering

  \resizebox{1.0\columnwidth}{!}{\includegraphics{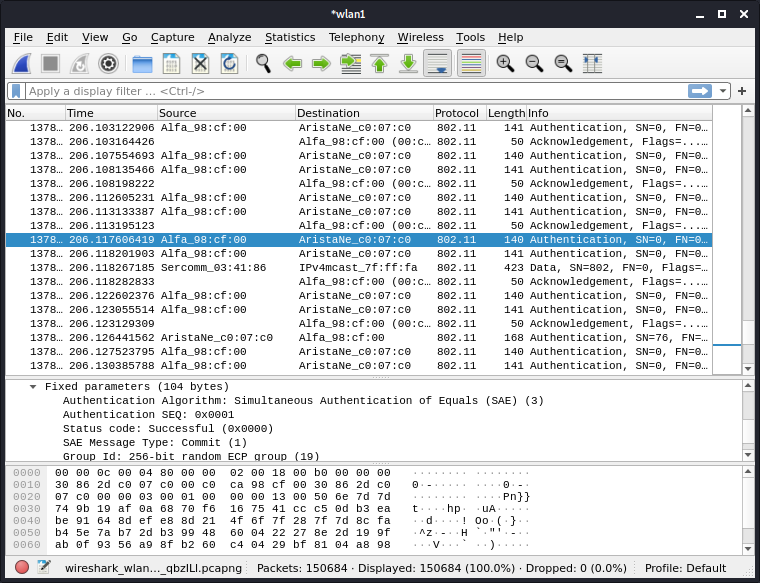}}
    %\caption{$W_{e}=3$,$W_{p}=2$,$W_{n}=1$}
%    \label{Avg_Tp_Diff_Wt} 
\end{minipage}
    \caption{(a) Terminal running the attack  (b) Wireshark snippet during the attack - flood of authentication requests}
   \label{authattack}
%    \label{AT}
\end{figure}
}

\subsubsection{Try to make a client use WPA2 instead of WPA3}\label{foolclient}\hfill\\
\textbf{Source:} This attack is from~\cite{depriv}.\\
\textbf{Impact:} Causes DoS-- prevents clients from connecting to the network. \\
\textbf{Our test result:} AP vulnerable and IDS does not detect it.\\
%Info about RSNE and EAPOL added earlier after Figure 1.
\textbf{Short Description:} A client aborts the EAPOL 4-way handshake if it finds any mismatch in the RSNE of the AP as seen in EAPOL Key message 3 versus what was advertised in the beacons. An attacker can exploit this connection abortion behavior. When the legitimate AP is configured to support WPA3 only, the attacker spoofs beacon frames advertising that the AP supports WPA2 only. By broadcasting such spoofed beacons at a rate higher than that of the legitimate AP, it can trick the client into connecting to the AP using WPA2 authentication. Later in message 3 of the 4-way handshake, the client realises that the AP supports WPA3 only and aborts the handshake. This causes DoS for any client trying to connect to the network. This attack is explained in detail in~\cite{depriv}.\\
\textbf{How to launch:} We set up and configure hostapd-2.9 \cite{hostapd} on our attack node, and make it run as an AP on the same channel and with the same SSID and MAC address as our target AP. We also decrease the beacon interval of our attacker AP to 16 ms (as compared to 100 ms set on the target AP) and set the key management to WPA2-PSK security (as compared to WPA3-SAE on the target AP).

\comment{
\begin{table}[h]
\begin{tabular}{ |l  l| } 
 \hline
 interface=wlan0 & \#Choose the interface on which we want to run hostapd\\
ssid=WPA3-Network & \#Set the SSID name\\
channel=1 & \#Set the channel\\
beacon\_int=16 & \#Set the beacon interval time in ms (default = 100)\\
bssid=30:86:2D:C0:07:C0 & \#Set the BSSID of the target AP\\
macaddr\_acl=0 & \#Set mac address filtering to off\\
wpa=1 & \#Set WPA2 standard. Set as '2' for WPA3\\
wpa\_passphrase=abcdefgh & \#Set the network password(unimportant, set to any value)\\
wpa\_key\_mgmt=WPA-PSK & \#Set the authentication type to use. 'WPA-PSK' or 'SAE'\\
rsn\_pairwise=CCMP & \#Set the encyrption algo to use - CCMP\\
ieee80211w=2 & \#Set management frame protection as mandatory\\
 \hline
\end{tabular}
\caption{\label{tab:2}WPA2 Config file}
\end{table}
}
\comment{
\begin{figure}[h]
    \centering
    \includegraphics[scale = 0.3]{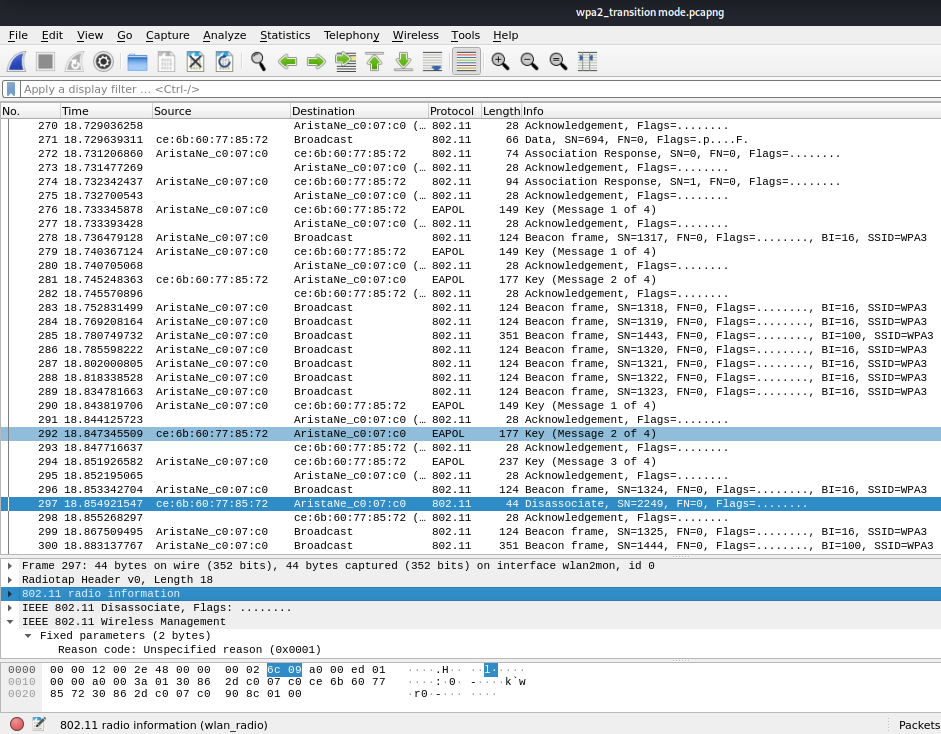}
    \caption{Client aborts the handshake after receiving message 3 when it realizes that AP indeed supports WPA3 only. The Beacons of length 124 with BI = 16 are from the attacker node and beacons of length 351 and BI = 100 are from our own AP.}
    \label{wpa2down}
\end{figure}
}
\subsubsection{Downgrade to WPA2 Attack (when AP is in Transition Mode)}\label{downgrade}\hfill\\
\textbf{Source:} This attack is from~\cite{dragon}.\\
\textbf{Impact:} It causes clients to connect to the network using the less secure WPA2 instead of WPA3. This makes the AP vulnerable to some of the attacks that were possible with WPA2.\\
\textbf{Our test result:} AP vulnerable and IDS does not detect it.\\
\textbf{Short Description:} WPA3 allows an AP to be in transition mode where it advertises the same SSID as supporting both WPA2 and WPA3. Similar to the above attack in~\Sectref{foolclient}, an attacker sends spoofed beacon frames at a high rate and makes the client infer that the AP supports WPA2 only. The client will then try to connect to the network using the weaker WPA2. Now while verifying the RSNE sent in message 3 of the handshake, the client sees that the  AP does support WPA2 and hence allows it to connect to the network using WPA2. Now it is exposed to all the attacks corresponding to WPA2.
This attack is explained in detail in~\cite{dragon}.
\\
\textbf{How to launch:} This attack can be launched in exactly the same way as the previous attack described in~\Sectref{foolclient}. We just set our target AP to be running in transition mode.   
%\textbf{Suggested countermeasures:}\\

\subsubsection{SAE Commit Values out of Range Attack}\hfill\\
\textbf{Source:} This attack is from~\cite{depriv}.\\
\textbf{Impact:} Causes DoS-- prevents clients from connecting to the network. \\
\textbf{Our test result:} AP vulnerable and IDS does not detect it.\\
\textbf{Short Description:} Once the client sends the first SAE commit message (see \Figref{SAE_auth}), the attacker can spoof the AP MAC address and in a race condition, reply with a ``commit value out of range rejection'' message before the legitimate AP replies. The client will abort the handshake on receiving the rejection message. Repeatedly sending such rejection messages before the legitimate AP causes DoS for the client. This attack is explained in detail in~\cite{depriv}.\\
\textbf{How to launch:} We modify the source code of hostapd-2.9 in such a way that the attacker's AP replies to the supplicant's commit message with a commit reply that contains the rejection status code 0x0001-- ``Unspecified failure''.
We found that our attacker was able to send the rejection message before the legitimate AP's reply in every test conducted and successfully cause DoS to the clients.

\comment{
\begin{figure}[h]
    \centering
    \includegraphics[scale = 0.3]{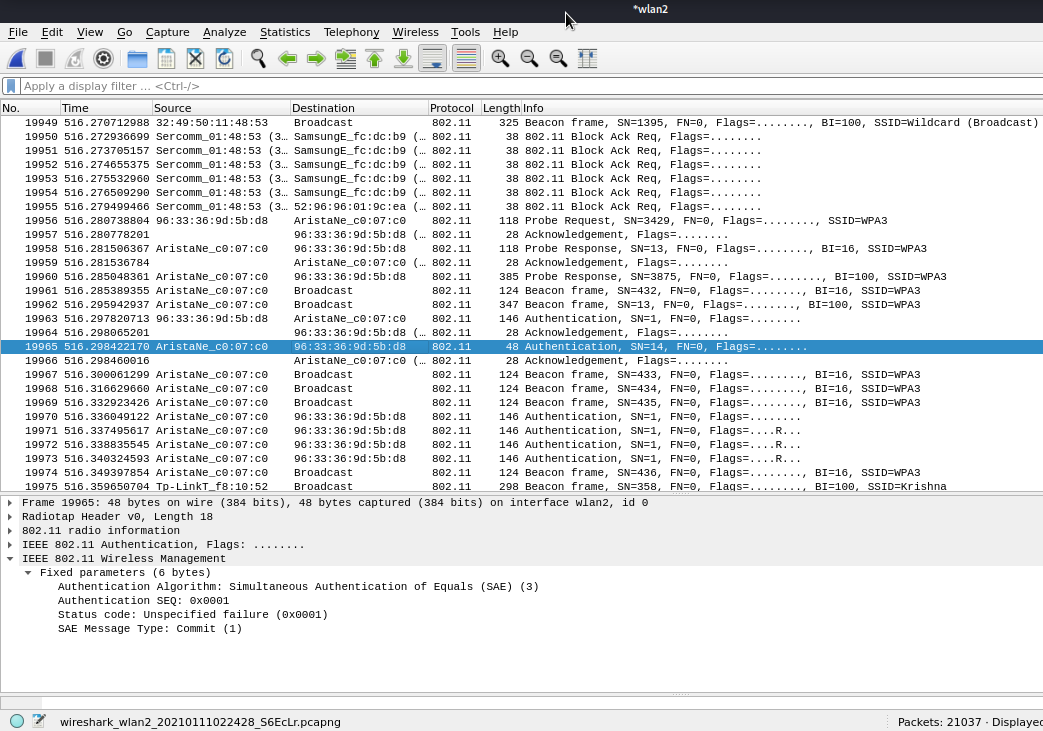}
    \caption{Packet 19965 is the `Unspecified failure' rejection message sent by the attacker. Client aborts the handshake on receiving this and does not reply to the  successful reply of the AP (Packet 19970).}
    \label{commit_attack}
\end{figure}
}
\subsubsection{SAE Unsupported Group Attack}\label{unsup}\hfill\\
\textbf{Source:} This attack is from~\cite{depriv}.\\
\textbf{Impact:} Causes DoS-- prevents clients from connecting to the network. \\
\textbf{Our test result:} AP vulnerable and IDS does not detect it.\\
\textbf{Short Description:} Once the client sends the first SAE commit message using a particular cryptographic group, similar to the commit out of range attack, an attacker can reply with a ``group unsupported'' message. The client will then abort the handshake and will have to resend its commit message using a different group. The attacker can again reply with a ``group unsupported'' message and keep doing this repeatedly causing DoS to the client. This attack is explained in detail in~\cite{depriv}.\\
\comment{
\begin{figure}[h]
    \centering
    \includegraphics[scale = 0.35]{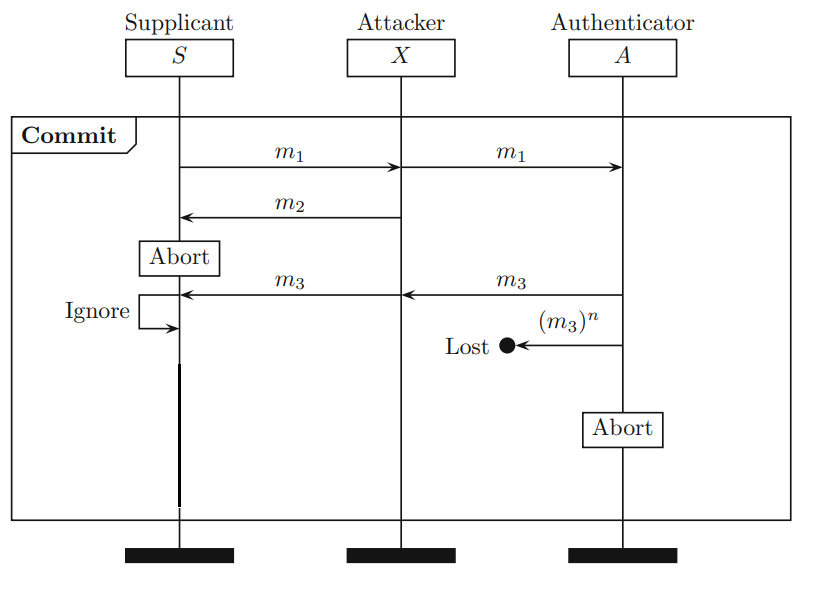}
    \caption{Here, m1 = commit message sent by client, m3 = commit reply by the legitimate AP and m2 = commit reply by the attacker which contains \textbf{group unsupported or commit value out of range} error code, hence rejecting the client request. (Figure source~\cite{depriv})}
    \label{commit_group}
\end{figure}
}
\textbf{How to launch:} We again modify the source code of hostapd-2.9 in such a way that the attacker's AP replies to commit messages with a rejection message that contains the status code 0x004d--  ``Authentication is rejected because the offered finite cyclic group is not supported''. We found that again 
our attacker was able to send the rejection message before the legitimate AP's reply in every test conducted and successfully cause DoS to the clients.
\comment{
\begin{figure}[h]
    \centering
    \includegraphics[scale = 0.3]{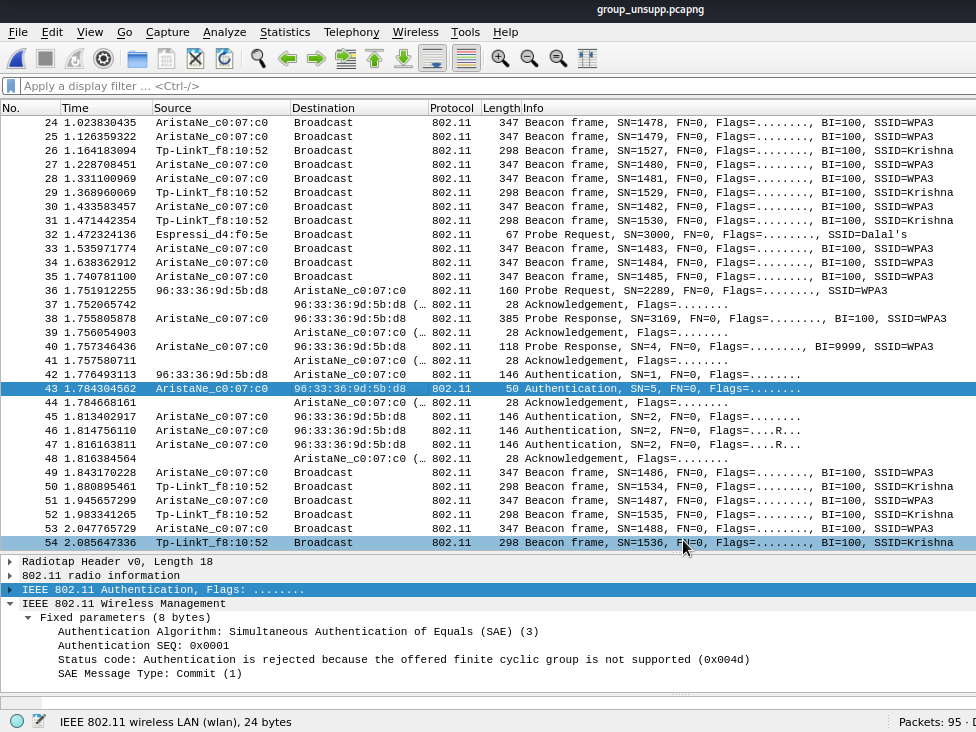}
    \caption{Packet 43 is the `Unsupported Group Used' rejection message sent by the attacker. Client aborts the handshake on receiving this and does not reply to the  successful reply of the AP (Packet 45).}
    \label{unsupp_attack}
\end{figure}
}

\subsubsection{Downgrade Group Attack}\hfill\\
\textbf{Source:} This attack is from~\cite{dragon}.\\
\textbf{Impact:} It makes a connection less secure by making use of a weaker cryptographic group in the SAE handshake. \\
\textbf{Our test result:} AP vulnerable and IDS does not detect it.\\
\textbf{Short Description:} The SAE mechanism can be carried out with a range of  different cryptographic groups. For this attack, rather than repeatedly replying to all client commit messages with ``group unsupported'' error codes, the attacker replies to only those commit messages that use strong groups and allows only weaker groups to pass. Hence, after starting with the strongest group, which gets rejected, the client eventually ends up using a weaker group for establishing its connection. This attack is explained in detail in~\cite{dragon}.\\
\textbf{How to launch:} This attack can be launched in a similar way as the previous one (\Sectref{unsup}). The only difference is that we need to modify the source code such that the rejection message is sent only when certain strong groups are being used, while allowing requests using weaker groups to pass.
\\  
%\textbf{Suggested countermeasures:}\\

\subsubsection{Timing Side Channel Attack}\label{timin}\hfill\\
\textbf{Source:} This attack is from~\cite{dragon}.\\
\textbf{Impact:} It leaks information about the password. If the password is weak, it is possible for an attacker to recover it. \\
\textbf{Our test result:} AP not vulnerable and IDS doesn't detect it.\\
\textbf{Short Description:} The time taken by an AP to respond to the authentication commit message sent by a client is a function of the password of the network and the MAC address of the client. This exposes it to a side channel timing attack, in which the attacker sends numerous first messages to the AP and records the average time it takes for the AP to respond. This timing information reveals certain information about the password.
An AP is vulnerable to this attack only if it uses the weak groups 22-24 (MODP groups) and 27-30 (Brainpool groups) during SAE authentication~\cite{802112016}. Our AP does not support these groups and hence is not vulnerable to this attack. This attack is explained in detail in~\cite{dragon}.\\
\textbf{How to launch:} In~\cite{dragondrain}, a tool and detailed steps to perform this attack are provided. It is important to note that this tool only carries out the attack, but does not specify if the AP implementation is vulnerable to it or not. We set up and configured wpa\_supplicant-2.9 \cite{wpasupp} on our attack node to make it run as a WPA3 client. We sent several SAE Authentication requests to the target AP using different cryptographic groups from 22-24 and 27-30. Each time we received a response from the AP rejecting the request with the reason ``group unsupported'' (0x004d). Our AP did not support the use of any of these weak groups, and hence was not vulnerable to this timing attack.

\subsubsection{Deauthentication Attack}\label{deauth}
\begin{figure}[h]
  \centering  
  \begin{minipage}[b]{0.5\linewidth}
        \centering
        \resizebox{1.0\columnwidth}{!}{\includegraphics{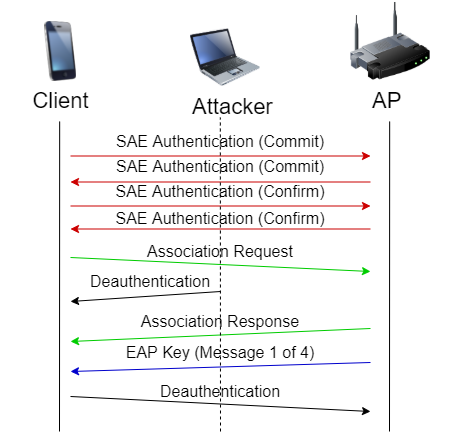}}
        \subcaption{}
 \end{minipage}%
 ~
\begin{minipage}[b]{0.5\linewidth}
  \centering

  \resizebox{1.0\columnwidth}{!}{\includegraphics{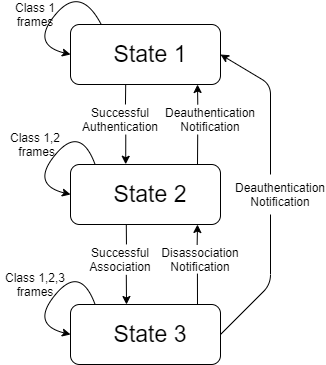}}
  \subcaption{}
\end{minipage}
    \caption{(a) The figure shows that by sending a deauthentication packet at the right time, an attacker can prevent new clients from joining the network. (b) The figure shows the state machine of a node running the 802.11 protocol during the establishment of a connection.}
\label{deauth_attack}
%    \label{AT}
\end{figure}

\begin{figure}[h]
    \centering
    \includegraphics[scale = 0.07]{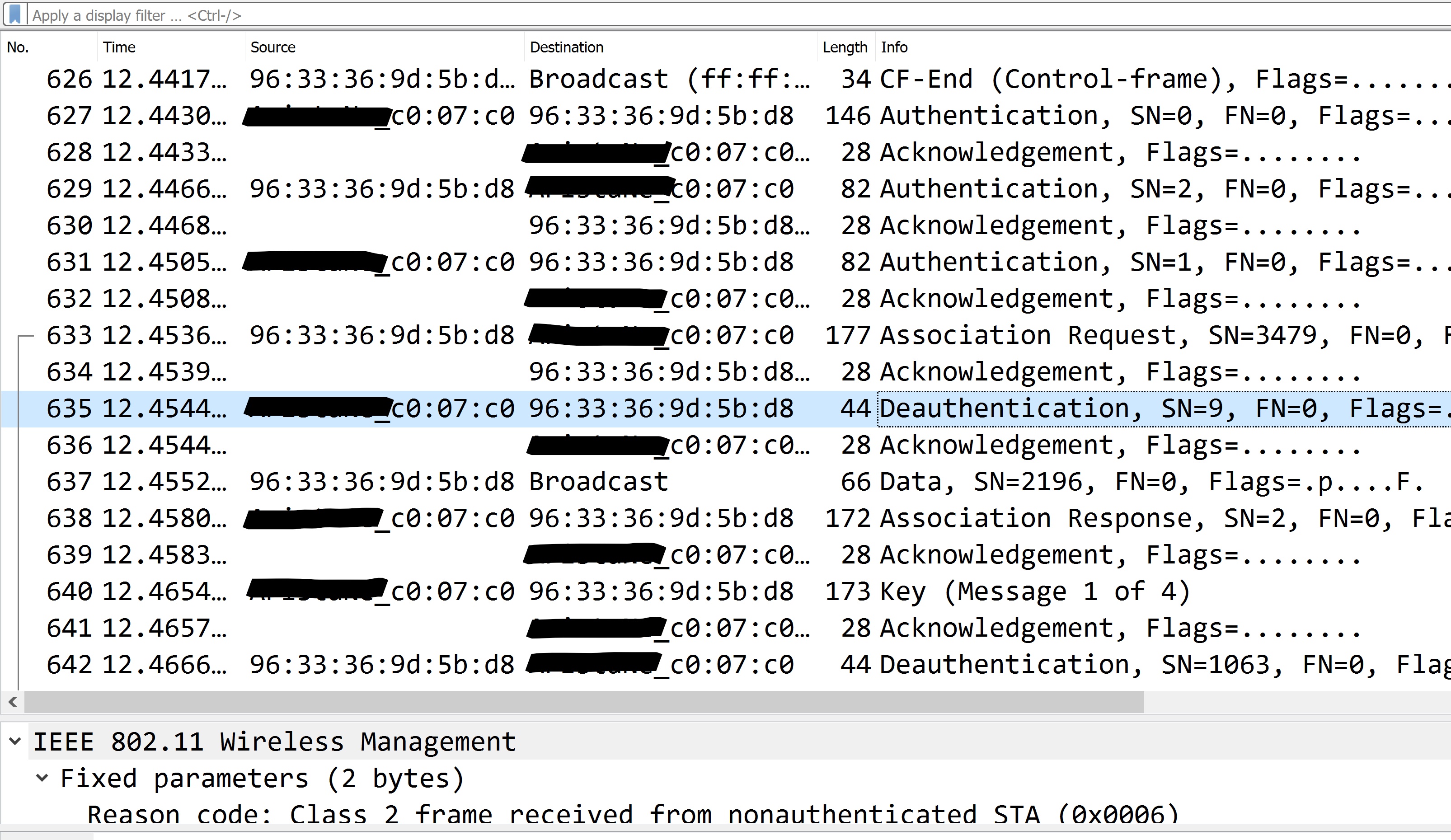}
    \caption{In this figure, the attacker AP, which spoofs the MAC address of the legitimate AP, sends a deauthentication packet (no. 635) to the client, which eventually leads to a mismatch of states between the client and legitimate AP and abortion of the handshake.}
    \label{deauthenti}
\end{figure}

\hfill\\
\textbf{Source:} This attack was possible with WPA2 \cite{deauthattack}. Below we show that a similar variant is still possible on a WPA3 network. \\
\textbf{Impact:} Causes DoS-- prevents clients from connecting to the network. \\
\textbf{Our test result:} AP vulnerable and IDS does not detect it.\\
\textbf{Short Description:} Management Frame Protection (MFP) introduced in 802.11w~\cite{80211w} and made mandatory with WPA3 prevents an attacker from spoofing a deauthentication frame after the 4-way handshake has taken place. However, an attacker can still spoof a deauthentication packet before this handshake and can cause DoS to a client trying to connect to the network. As shown in~\Figref{deauth_attack} (a), just after the client sends an association request to the AP, the attacker sends a deauthentication packet to the client. The client now goes back to State 1 of its state machine as shown in~\Figref{deauth_attack} (b) and expects frames only of type Class 1 as defined by the 802.11 protocol \cite{802112016}. The AP is unaware of this and continues to respond to the association request with an association response followed by the EAPOL message 1 of the 4-way handshake as it would normally do. The client in State 1 does not expect to receive these messages and replies to them with a deauthentication packet with reason code 7--  ``Class  3  frame  received  from  nonassociated STA''. The AP accepts this deauthentication packet and aborts the handshake. The attacker keeps doing this every time a client tries to establish a connection with the AP, thus causing DoS to the client.\\
\textbf{How to launch:} To implement this attack, we again modify the source code of hostapd-2.9 in a way that the attacker's AP sends a deauthentication packet to the client right after the client completes authentication and sends an association request to the legitimate AP. We found that our attacker was able to send the deauthentication packet before the legitimate AP's reply in every test conducted and successfully cause DoS to the clients. \Figref{deauthenti} shows the packets captured during the attack, which match with the packet sequence we expect theoretically from~\Figref{deauth_attack} (a).

\subsubsection{Beacon/ Probe frames Flood Attack}\label{beaconfld}
\begin{figure}[h]
  \centering  
  \begin{minipage}[b]{0.5\linewidth}
        \centering

        \resizebox{1.0\columnwidth}{!}{\includegraphics{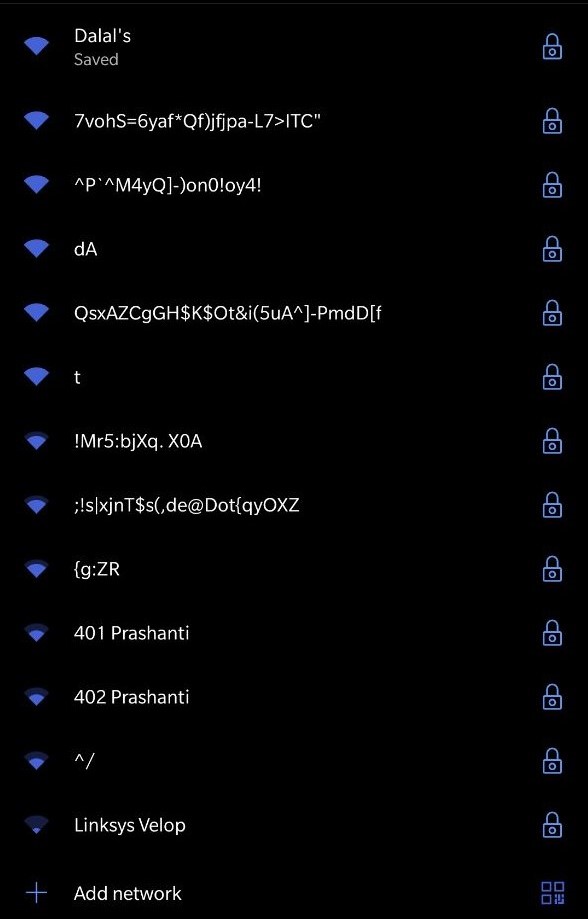}}
        \subcaption{}
         %\caption{$W_{e}=W_{p}=W_{n}=1$}
%    	\label{Avg_Tp_Same_Wt} 
 \end{minipage}%
 ~
\begin{minipage}[b]{0.5\linewidth}
  \centering

  \resizebox{1.0\columnwidth}{!}{\includegraphics{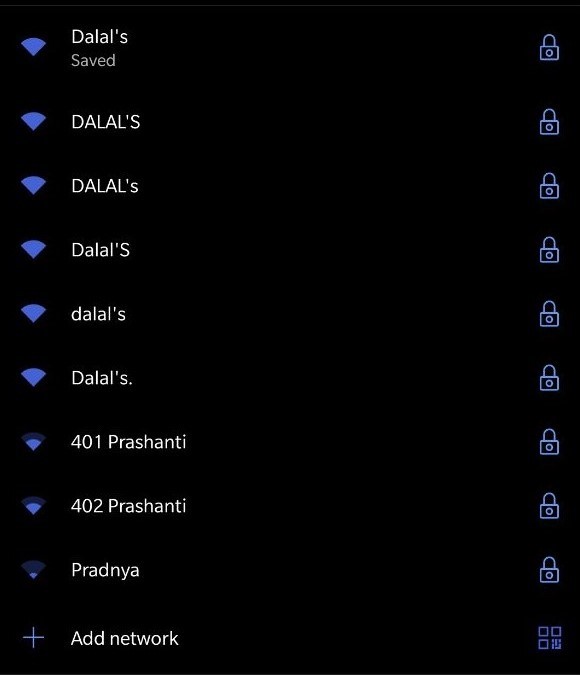}}
  \subcaption{}
    %\caption{$W_{e}=3$,$W_{p}=2$,$W_{n}=1$}
%    \label{Avg_Tp_Diff_Wt} 
\end{minipage}
    \caption{(a) Beacon flood with random SSIDs, (b) Beacon flood with confusing SSIDs}
\label{beaconflood}
%    \label{AT}
\end{figure}

\hfill\\
\textbf{Source:} This attack was possible with WPA2. In this paper, we show that it is still possible on a WPA3 network. \\
\textbf{Impact:} Confuses new clients trying to find the legitimate AP. \\
\textbf{Our test result:} AP vulnerable and IDS does not detect it.\\
\textbf{Short Description:} Since the beacons are not protected, an attacker can flood the network with beacons advertising similar or different SSIDs compared to the legitimate AP. Likewise, when a client sends a Probe request frame, the attacker can flood the network with Probe response frames from similar or different SSIDs. Similar SSIDs can confuse a user, while a huge list of different (say 200) SSIDs can make it time consuming for clients to find the correct one. This is illustrated in~\Figref{beaconflood}.\\
\textbf{How to launch:} This attack can be easily launched with the MDK3 tool~\cite{mdk3}, which comes pre-installed with Kali Linux.

\section{Signature based IDS techniques}\label{sigdet}
\subsection{High Level Design and Overview}
\begin{figure}[h]
    \centering
    \includegraphics[scale = 0.32]{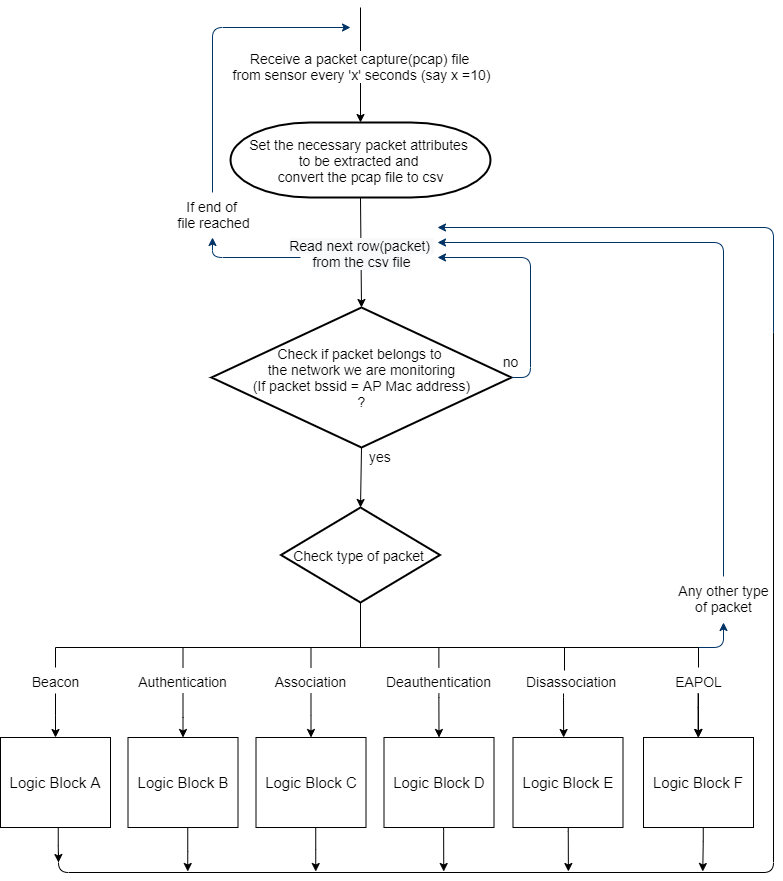}
    \caption{High level design of the IDS}
    \label{high_level}
\end{figure}

Our IDS sensor monitors and captures the packets being exchanged in the network into a `.pcap' file. The IDS then processes this `.pcap' file. It extracts the necessary attributes for each type of packet and then converts it into a `.csv' file. Then, each packet (row) is read from this `.csv' file one by one, and appropriate conclusions are drawn. Each of the logic blocks (A-F) in~\Figref{high_level} analyses the packet received, applies some logic, updates the state of the IDS, and if there is any attack detected, prints the same along with the packet timestamp. We look at how we implement each of these blocks in~\Sectref{detailed}. 

\subsection{Packet Attributes Extracted for Each Type of Packet}
\begin{itemize}
    \item \textbf{All frames in general}\hfill\\
    - Serial number of the frame (frame.number)\\
    - Timestamp of the frame (frame.time)\\
    - Source address (wlan.sa)\\
    - Receiver address (wlan.ra)\\
    - BSSID of the frame (wlan.bssid)\\
    - Sequence number (wlan.seq)\\
    - Type of frame (wlan.fc.type)\\
    - Subtype of frame (wlan.fc.subtype)
    \item \textbf{Beacon/ Probe frame}\hfill\\
    - Beacon interval (wlan.fixed.beacon)\\
    - Beacon timestamp (wlan.fixed.timestamp)\\
    - SSID advertised (wlan.ssid) \\
    - Number of authentication protocols\hfill\\ supported (wlan.rsn.akms.count)\\
    - Type of authentication protocol\hfill\\ supported (wlan.rsn.akms.type)
    \item \textbf{Authentication frame}\hfill\\
    - Type of authentication algorithm used \hfill\\(wlan.fixed.auth.alg)\\
    - Authentication sequence number\hfill\\ (wlan.fixed.auth\_seq)\\
    - Status code of the event (wlan.fixed.status\_code)\\
    - SAE message type (wlan.fixed.sae\_message\_type)\\
    - Cryptographic group used for\hfill\\ authentication (wlan.fixed.finite\_cyclic\_group)
    \item \textbf{Association frame}\hfill\\
    - SSID advertised (wlan.ssid) \\
    - Number of authentication protocols\hfill\\ supported (wlan.rsn.akms.count)\\
    - Type of authentication protocol\hfill\\ supported (wlan.rsn.akms.type)\\
    - Association ID (wlan.fixed.aid)
    \item \textbf{Deauthentication frame}\hfill\\
    - Reason for Deauthentication\hfill\\ (wlan.fixed.reason\_code)
    \item \textbf{Disassociation frame}\hfill\\
    - Reason for Disassociation (wlan.fixed.reason\_code)
    \item \textbf{EAPOL frame}\hfill\\
    - Key descriptor indicating EAPOL frame\hfill\\ (eapol.keydes.type)\\
    - EAPOL message number - 1, 2, 3 or 4\hfill\\ (wlan\_rsna\_eapol.keydes.msgnr)

\end{itemize}

\subsection{Detailed Design of the IDS to Detect Specific Attacks} \label{detailed}

\subsubsection{Design to Detect SAE Authentication Flood Attacks}
\label{SSSC:SAE:auth:flood:detection}
We keep track of the last eight authentication frames sent to the AP at any given point of time. That is, we maintain a buffer of size eight containing the timestamps of the last eight authentication frames exchanged in the network (and so it will be updated using the FIFO rule when a new authentication frame is received). If at any point, the time difference between the first and last frames of the buffer is less than 500 ms, then we note that an abnormal event has occurred. A rate of 8 frames per 500 ms (= 16 frames/ sec) is actually low and an attacker would always have to send packets at a higher rate for the attack to be successful~\cite{dragon}. But keeping our detection threshold at this low rate helps increase the detection accuracy. Then to lower the false positive rate, we wait to see if such an abnormal event is detected again at least 10 times every minute for the next 3 minutes. If it is, only then we say that an authentication flood attack is detected. (We use the fact that the attacker would have to keep flooding the network continuously to successfully cause a DoS attack. On the other hand, in any normal scenario, there would not be such a high rate of authentication frames being consistently present in the network for such a long time. Even in a case where an AP restarts and all its clients try to connect to the network simultaneously, they would end up doing so within the first minute itself. Hence this case will not be falsely detected as an attack.)

\begin{remark}
1) We only count the authentication frames sent towards the AP and not the ones sent by the AP as replies.\\
2) We also do not count the second authentication frame (confirm message) sent by the client. So effectively, we are only keeping track of the number of connection requests received by the AP.\\
3) We also incorporate an additional check to reduce false positives. When an abnormal event is detected, we store the eight MAC addresses of the clients from which the eight connection requests were received. Then we wait for a short time and see if the majority of the connections from those eight addresses turn out to be successful. If they do, then we do not consider that as an abnormal event. We use the fact that only legitimate clients, having possession of the network password, would be able to establish a successful authentication. In other words, we only count the unsuccessful authentication connection requests and ignore the successful ones. We consider a connection successful if the AP replies to the client with a confirm message containing the status code as successful.
\end{remark}

We add the above detection logic to block B (see~\Figref{high_level}). 

\subsubsection{Design to Detect WPA2 Downgrade Attacks}
We keep track of the last two beacon frames having the same SSID and BSSID as that of our AP. In particular, we maintain a FIFO buffer of size two containing all the necessary information of the last two beacons. If at any time we see that there is a mismatch in the RSNE information of these two beacon frames, then we note that an abnormal event has occurred. In particular, if between the two beacons the number of authentication protocols (wlan.rsn.akms.count) differs or the type of authentication protocols supported (wlan.rsn.akms.type) changes from WPA3 to anything lower, then that is an abnormal event. If such an abnormal event is detected at least 4 times in a span of 5 seconds, then we conclude that a WPA2 downgrade attack is being executed. This logic is able to detect the attacks in both Sections~\ref{foolclient} and~\ref{downgrade}. 

\begin{remark}
1) APs generally have a beacon interval set to 102.4 ms. Even if an AP sets the beacon interval to as high as 1024 ms, it will still broadcast 4 beacons in any 5-second window. Hence still, 4 abnormal events will be raised in a 5-second window, and our IDS will be able to detect the attack successfully.\\
2) If the time difference between the last two beacons is greater than 10 seconds, then we do not check for the abnormal event condition; we ignore it. This is done to avoid a false positive resulting from the restarting or reconfiguration of the AP. 
However, if a secure communication channel between the AP and the IDS is available, then when the AP has restarted, an `AP restarted' signal can be shared with the IDS by the AP. We can then use this signal to conclude that the AP has restarted and possibly reconfigured. In such a case, we discard any abnormal events detected just before and just after such a signal is received. This would again help in avoiding any false positives.\\
3) When our AP was configured to run in Transition Mode (supporting both WPA2 and WPA3), it advertised itself using just one beacon, and not two separate beacons, per beacon interval. The RSNE of the beacon carried information that the AP supported both SAE and PSK type of authentication.
\end{remark}

We add the above detection logic to block A (see~\Figref{high_level}). 

\subsubsection{Design to Detect Deauthentication Attacks}
Once we detect that a deauthentication packet is exchanged between a client (victim) and our AP, there should be no packet of type Association or EAPOL being exchanged between these two parties for the next short interval of time (say 3 seconds). If such a packet is detected within the short interval, then we conclude that an attack is being executed.

It can happen that the deauthentication packet sent by the attacker to the victim client is not sniffed by our IDS sensor, which is located close to the AP and far from the client. In this case, the attack will pass undetected. So we incorporate an additional signature to detect this attack. We look for the condition where a client has sent an association request, the AP has replied with a successful association response and then the client has sent a deauthentication packet to the AP with reason code 7-- ``Class 3 frame received from nonassociated STA''. When we detect such a sequence of packets being exchanged, then we say that an abnormal event has happened. This is because once a client has sent an association request, it cannot reply to a successful association response with a deauthentication packet with reason code 7. This can only take place if before receiving the association response, the client itself was deauthenticated and is now expecting frames of only Class 1 type as defined by the 802.11 protocol. If such an abnormal event repeats in a short time interval, then we conclude that a deauthentication attack is taking place.\\
We add this detection logic to blocks D, C, and F.\\
\textbf{Logic Block D (Deauthentication frame)}\\
Store the victim client's MAC address (say in `deauth\_client' array) and the time at which the deauthentication packet is seen (say in `deauth\_time' array).\\
\textbf{Logic Block C (Association frame)}\\
Check if source or destination address of the current packet lies in the `deauth\_client' array. If it does not, then do nothing and move on to read the next packet. If it does, then note down the index at which that address is present in `deauth\_client'. Check the time difference between the time the association packet is received and the corresponding time stored at the same index in `deauth\_time' array. If this time difference is less than 3 seconds, then count it as an abnormal event.\\
\textbf{Logic Block F (EAPOL frame)}\\
Same as logic block C.

\subsubsection{Design to Detect Group Downgrade and Commit Value out of Range Attacks}
At any instant if we see that an authentication rejection packet is sent by the AP to a particular client, we then monitor the channel for the next 500 ms to see if a successful authentication packet is sent by the AP to the same client. If such a packet is found, then we conclude that the group unsupported or commit value out of range attack is being performed. The authentication rejection packet should have status code ``group unsupported'' (0x004d) or ``commit value out of range'' (0x0001) and the successful authentication packet should have status code ``successful'' (0x0000).

\begin{remark}
It can happen that the authentication packet sent by the attacker (masquerading as AP) to the victim client is not sniffed by our IDS sensor, which is located at the AP, far from the client. In this case, the attack will pass undetected. One way to solve this issue is by having multiple sensors at different locations, so that every packet is sniffed by at least one IDS.
\end{remark}

We add the above detection logic to block B (see~\Figref{high_level}). 

\subsubsection{Design to Detect Timing Side-Channel Attacks}
Similar to the detection of the SAE Authentication flood attack (see Section~\ref{SSSC:SAE:auth:flood:detection}), we keep a count of the number of unsuccessful authentication requests received by the AP. In this case, these requests may not be flooded in a short time period but could be spread out across a longer duration. Hence, we do not use the request transmission times, and, instead, conclude that a timing side channel attack is being performed when the detected count of these unsuccessful authentication requests reaches a very high value, say 500. 

In order to not have false positives, this count value is reset to zero every 24 or 48 hours. Also, it should be reset to zero after an authentication flood attack is detected (as the signature we defined for the authentication flood attack is also a signature for the timing attack).

We add the above detection logic to block B (see~\Figref{high_level}). 

\subsubsection{Design to Detect Beacon Flood Attacks}
When the IDS is first booted, we put it into a learning phase where it stores the MAC addresses and SSIDs of the APs in its neighborhood. Thus it builds and maintains an authorised APs list. After a short period (say 2-3 minutes), we change the IDS mode into detection phase where it monitors the network and marks an abnormal event when it receives beacons for which either MAC address or SSID does not lie in the authorized APs list. When a number of such abnormal events are detected in a short time interval, then we conclude that a beacon flood attack is in progress. Specifically, we draw this conclusion when 5 such different abnormal events occur in a span of 10 seconds.
The manufacturer of our AP provides a Wireless Cloud Manager server that stores the current state of all our APs. When an AP restarts or a new AP is added to the network, this information gets updated on the server. At the IDS, using the Application Programming Interfaces (APIs) provided by the manufacturer of the AP, we poll our APs' Wireless Manager server to check for this information. Whenever we see that an AP is restarted or a new AP is added to the network recently, we put the IDS back into the learning phase for a short period. This helps in reducing false positives. 

We add the above detection logic to block A (see~\Figref{high_level}). We can use this same logic with probe frames instead of beacons to detect Probe frames Flood attack.

\begin{remark}
It is important to note that the particular values and thresholds chosen in the design of our IDS are based on our experimental observations. They can be easily changed if found sub-optimal.
\end{remark}

\section{Mitigation of Impact of Detected Attacks}\label{next}
Except for the SAE flood and timing side-channel attacks (Sections~\ref{auth} and~\ref{timin}), in all the other attacks described in Section~\ref{list_attack}, the attacker sends some malicious packets to the clients. So once we detect any of these attacks, we record the MAC addresses of the affected clients. We then send them a notification. These affected clients can later take appropriate actions from their side.

Also, we can track down any rogue transmitters in the vicinity. We do this by maintaining a list of authorized APs. Using the APIs provided by the manufacturer of our AP, we poll our APs' Network Management System (NMS) to get information about the authorized APs set up in our location and maintain this list. If there is any AP not present in this list but present in the vicinity, then it is marked as rogue and may be banned.

Also, if we have multiple IDSs set up in a region, then once an attack is detected, we can find the approximate physical location of the attacker. We do this by first identifying which packets are sent by the attacker. We mark those packets and measure their received signal strength (RSSI value) at each of the IDSs. We then try to find the location of the transmitter of those packets using signal triangulation~\cite{nist}.
 
\section{Experimental Results}\label{exp}
\subsection{Setup and Configuration}
We use a laptop running Kali Linux v2020.4 as our attack node.
Along with it, we use a Network Interface Card (NIC) that monitors the channel, injects packets as required and supports the 802.11ac protocol. Some of the attacks also require this NIC to be using an Atheros-based chipset. Hence we use the popular Alfa AWUS036NHA based on the Atheros AR9271 Chipset. We have used this to perform all the attacks described in this paper.

Additionally, we also use another NIC to sniff, collect and log packets from the channel while the attacks are being performed. The only requirement for this NIC is that it should support monitor mode. We use the Leoxsys LEO-HG150N NIC, but any other cheap NIC with monitor mode support can be used. Alternatively, the inbuilt NIC of a laptop could also be used if it supports monitor mode.

Lastly, we use an enterprise AP against which we launch our attacks. It has support for the 802.11ax protocol, WPA3 Wi-Fi security and has three radios. We set the first radio to run as a WPA3 AP on the frequency 2.4 GHz; we keep the second radio unused and set the third radio to run as the IDS offered by the manufacturer. (It is important not to set the second radio to run as a WPA3 AP on the frequency 5 GHz as this will lead to confusion while performing some of the attacks; note that all the attacks we perform are only on the 2.4 GHz channel). For the client, we use a mobile phone (OnePlus7) running Android 11 and a Dell laptop running Windows 10, both supporting WPA3.

\begin{figure}[h]
    \centering
    \includegraphics[scale = 0.32]{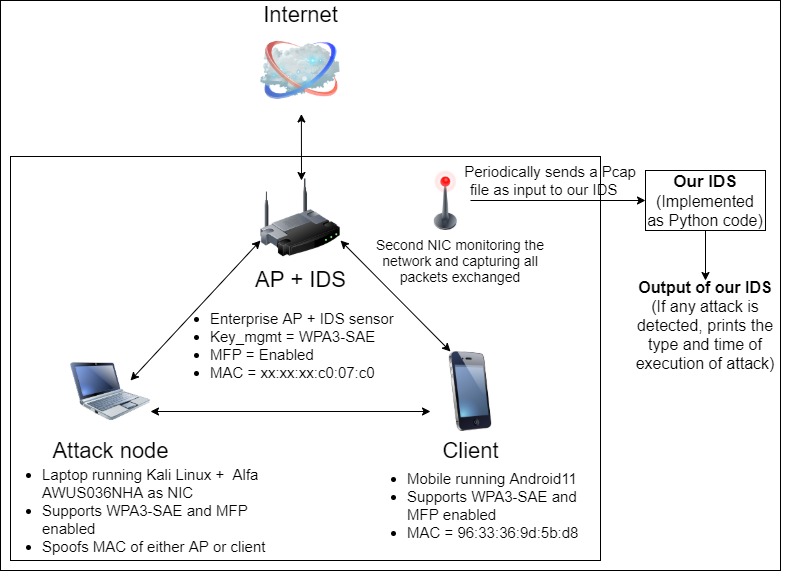}
    \caption{Experimental setup}
    \label{testbed}
\end{figure}

~\Figref{testbed} depicts our experimental setup.

\subsection{Attack Execution and Detection by IDS}
We perform attacks on the AP from our attack node and capture the packets exchanged during each attack in a `.pcap' file. This file is fed as an input to our IDS logic, which is coded in Python. If there is any attack detected in some part of the packet capture, the IDS prints the type and time of execution of the attack. We do this for each of the nine attacks described in~\Sectref{list_attack}.
 
\begin{figure}[h]
    \centering
    \includegraphics[scale = 0.07]{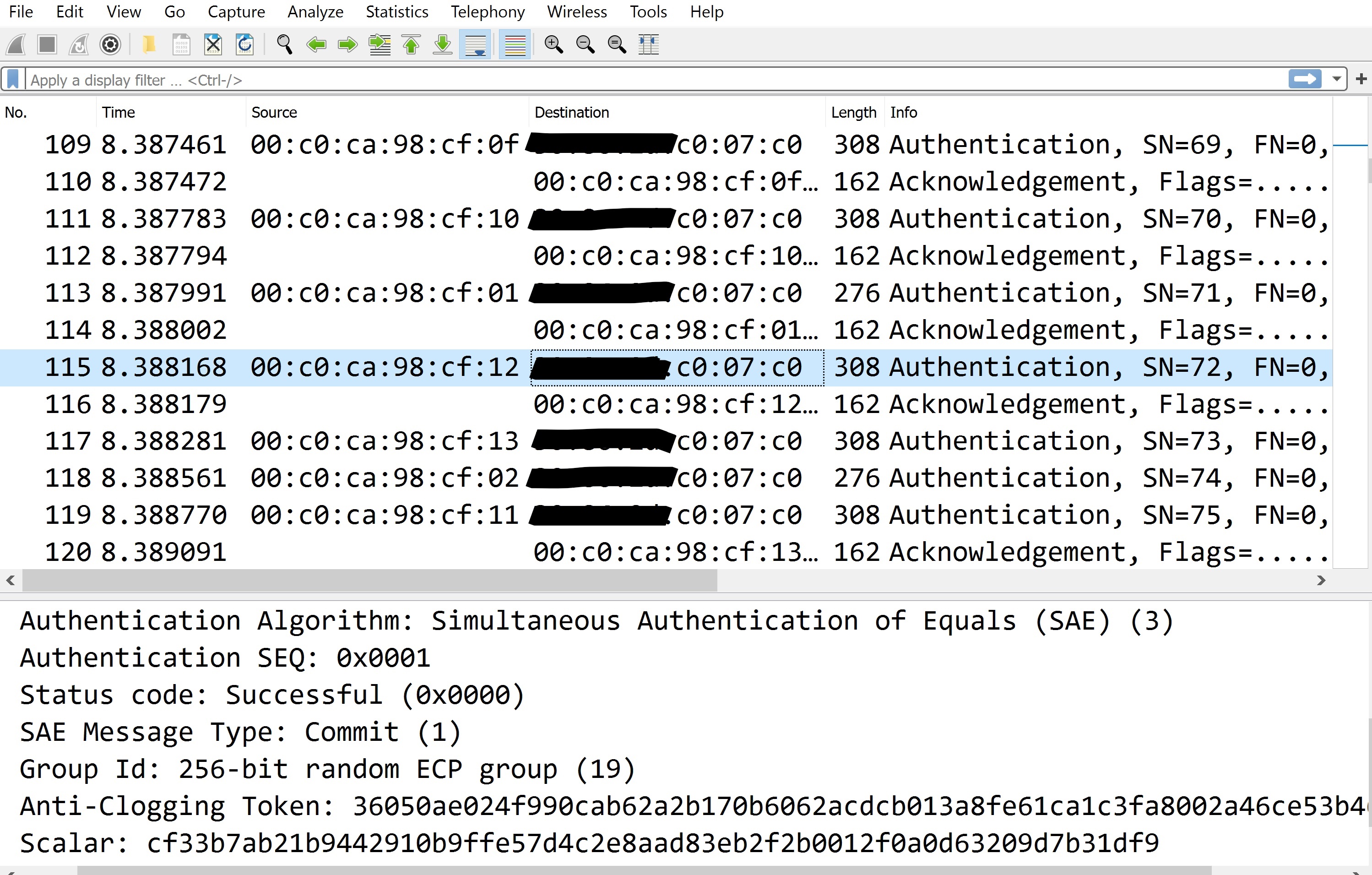}
    \caption{The figure shows the packet capture of the SAE authentication flood attack. We can see that seven authentication frames are received by the AP in 1.3 ms.}
    \label{authenti}
\end{figure}

\begin{figure}
\centering
\begin{subfigure}{0.5\textwidth}
   \includegraphics[width=1\linewidth]{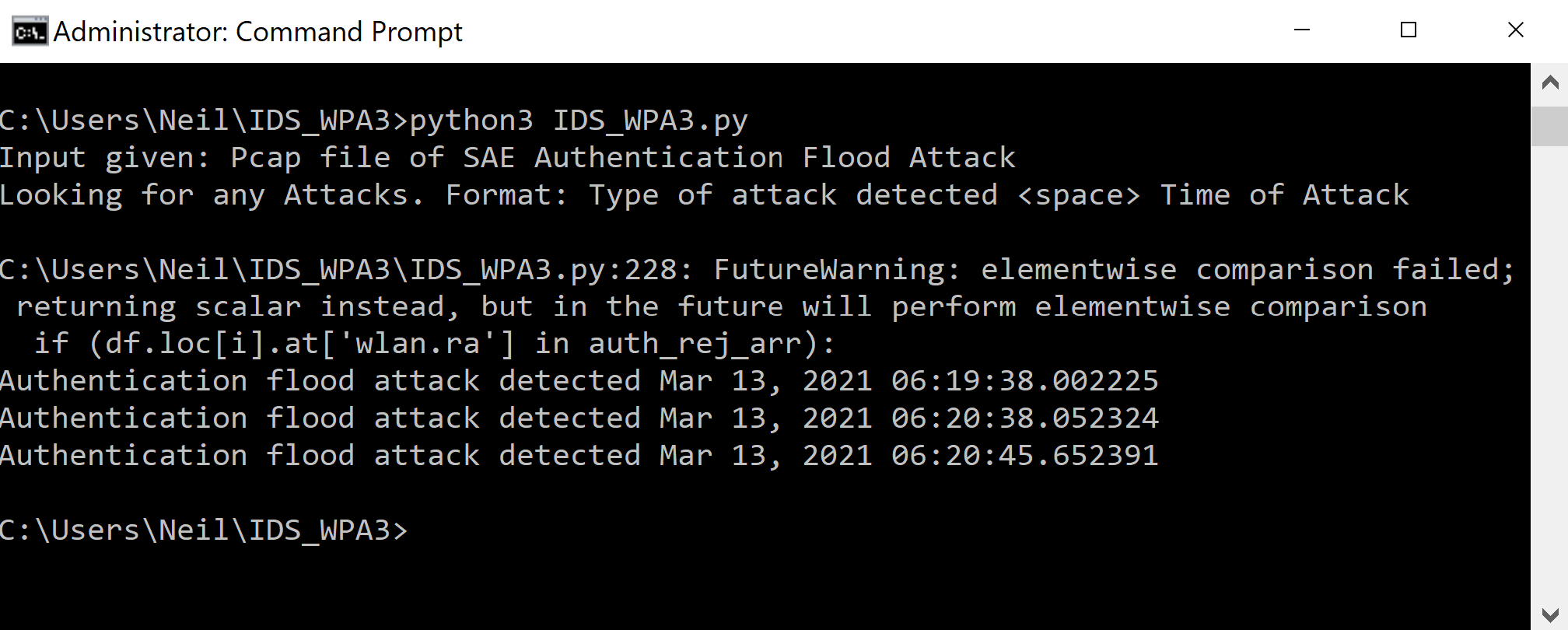}
   \subcaption{}
\end{subfigure}

\begin{subfigure}{0.5\textwidth}
   \includegraphics[width=1\linewidth]{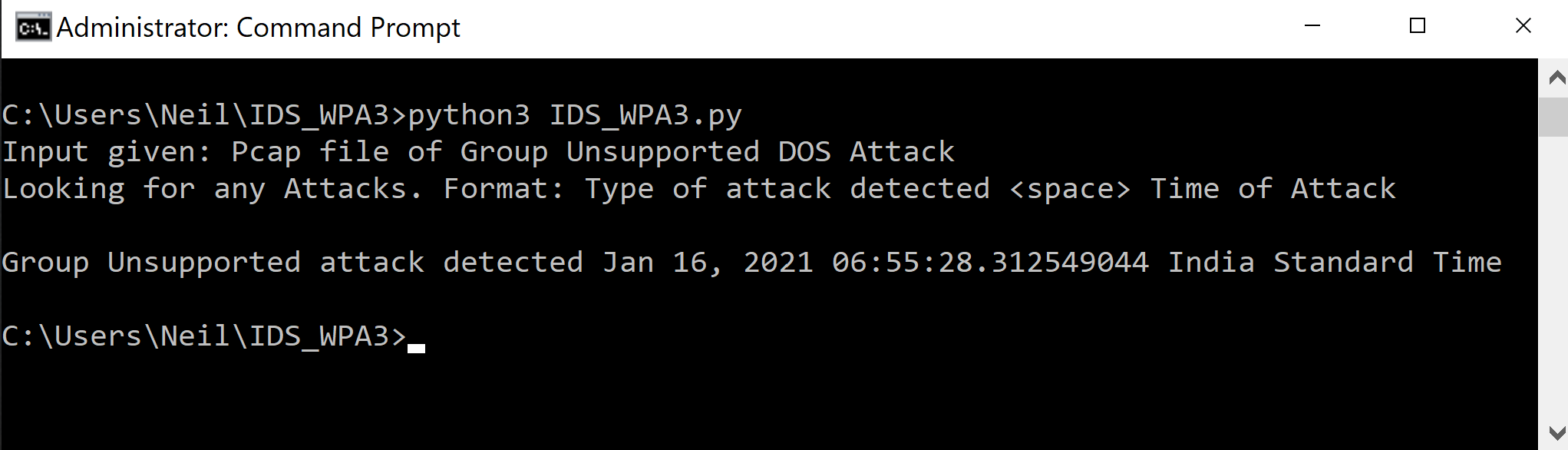}
   \subcaption{}
\end{subfigure}

\caption{The figure shows the response of the IDS when the packet capture of (a) the SAE authentication flood attack, (b) the SAE unsupported group DoS attack, is given as input.}
\label{new}
\end{figure}

For example,~\Figref{authenti} shows the SAE authentication flood attack with an authentication flood rate of 200 frames/ sec.
 As seen in its packet capture, the AP receives a flood of authentication requests from our attack node.~\Figref{new} (a) shows the response of the IDS when this packet capture is fed as input. We see that the IDS is successfully able to detect this flood attack. We also re-perform the attack with the authentication flood rate reduced to as low as 20 frames/ sec. Even at this rate, our AP was still vulnerable. It either was unable to accept new connections or only accepted them after a considerable delay. For our IDS, we had set the detection threshold to 16 frames/ sec. Hence, as expected, our IDS was able to detect the attack even at the above low rate. 

\begin{figure}[h]
    \centering
    \includegraphics[scale = 0.067]{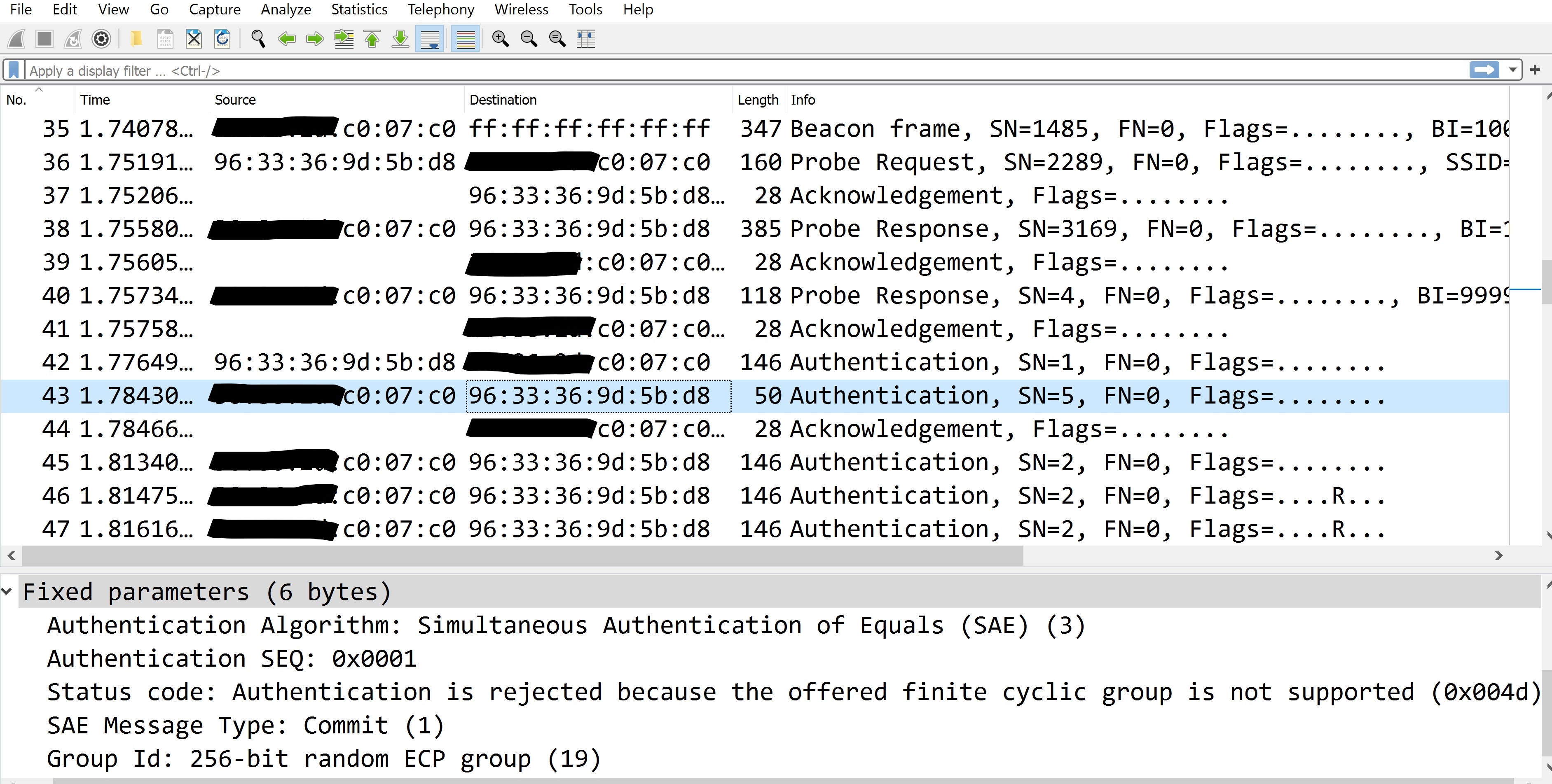}
    \caption{The figure shows the packet capture of the SAE unsupported group DoS attack. The highlighted packet 43 with status code 0x004d is sent by the attacker, which spoofs the MAC address of the AP.}
    \label{unsuppgroup}
\end{figure}

~\Figref{unsuppgroup} shows the SAE unsupported group DoS attack. We see that the packets we captured match our theoretical expectation of the attack packet sequence. Packet 42 is the first commit message sent by the client (96:33:36:9d:5b:d8) to the legitimate AP (xx:xx:xx:c0:07:c0). Packet 43 is the one sent by the attacker, which spoofs the MAC address of the legitimate AP. Its status code as seen in~\Figref{unsuppgroup} is set to 0x004d-- ``Authentication rejected because group is not supported''. Packet 45 is the successful commit reply sent by our legitimate AP. Packets 46 and 47 are the same as packet 45 but with the retry bit set to 1-- the AP re-sends its commit message since it receives no reply from the client after packet 45.~\Figref{new} (b) is our IDS response when this packet capture file is fed as input. We see that it correctly detects a Group Unsupported attack and the time of detection is the same as that of packet 45. We observed that when the attack was in operation, no new clients were able to join the network. Our clients would either get stuck on ``Obtaining IP Address" or get the error-- ``Check Password and Try Again" (even when we had entered the correct credentials). As soon as the attack was stopped, all clients were able to successfully connect to the network.

\comment{
\begin{figure}[t!]
  \centering  
  \begin{minipage}[b]{0.5\linewidth}
        \centering

        \resizebox{1.0\columnwidth}{!}{\includegraphics{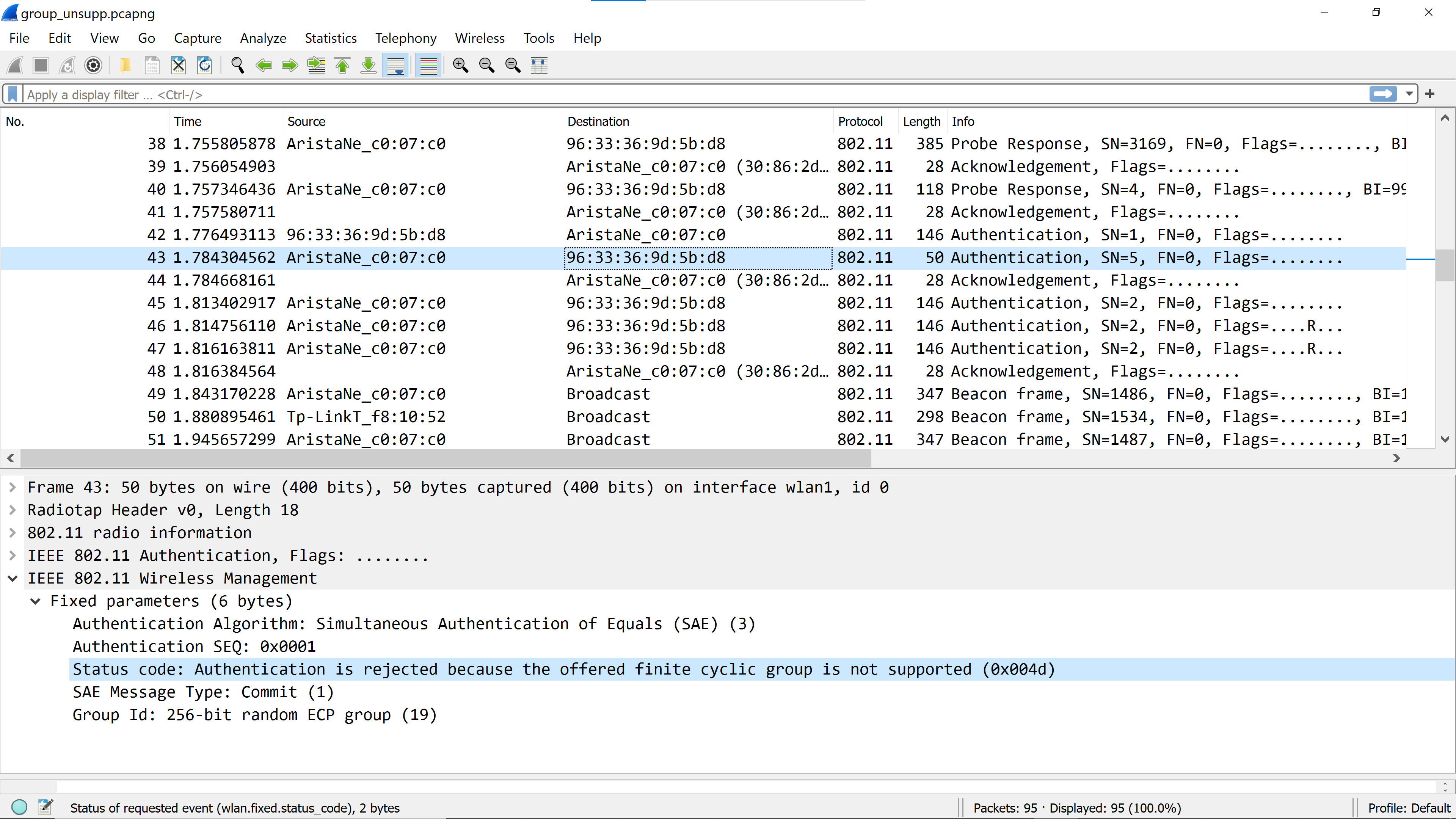}}
         %\caption{$W_{e}=W_{p}=W_{n}=1$}
%    	\label{Avg_Tp_Same_Wt} 
 \end{minipage}%
 ~
  \begin{minipage}[b]{0.5\linewidth}
        \centering

        \resizebox{1.0\columnwidth}{!}{\includegraphics{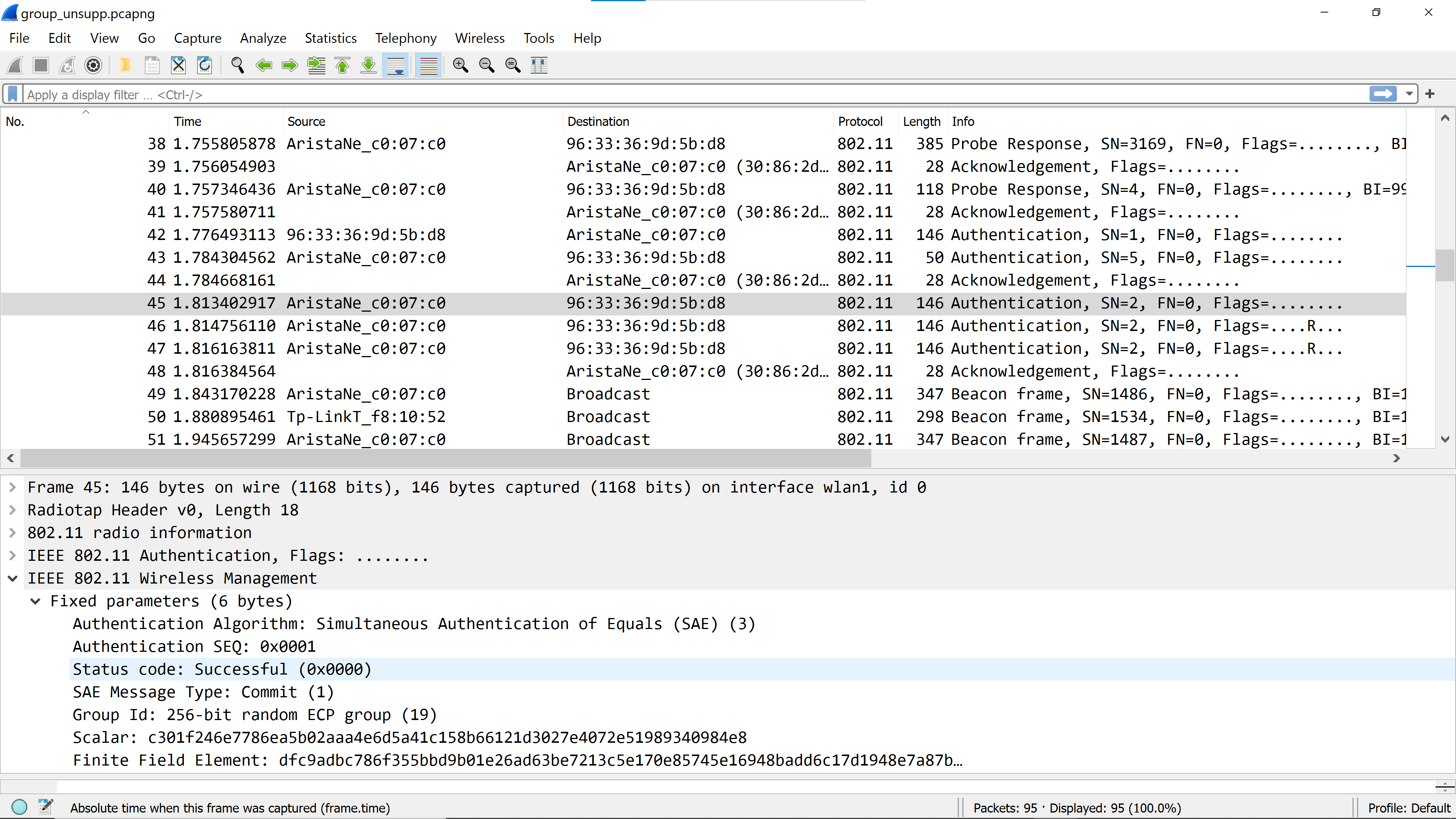}}
         %\caption{$W_{e}=W_{p}=W_{n}=1$}
%    	\label{Avg_Tp_Same_Wt} 
 \end{minipage}%
 ~
\newline
  \begin{minipage}[b]{0.5\linewidth}
        \centering

        \resizebox{1.0\columnwidth}{!}{\includegraphics{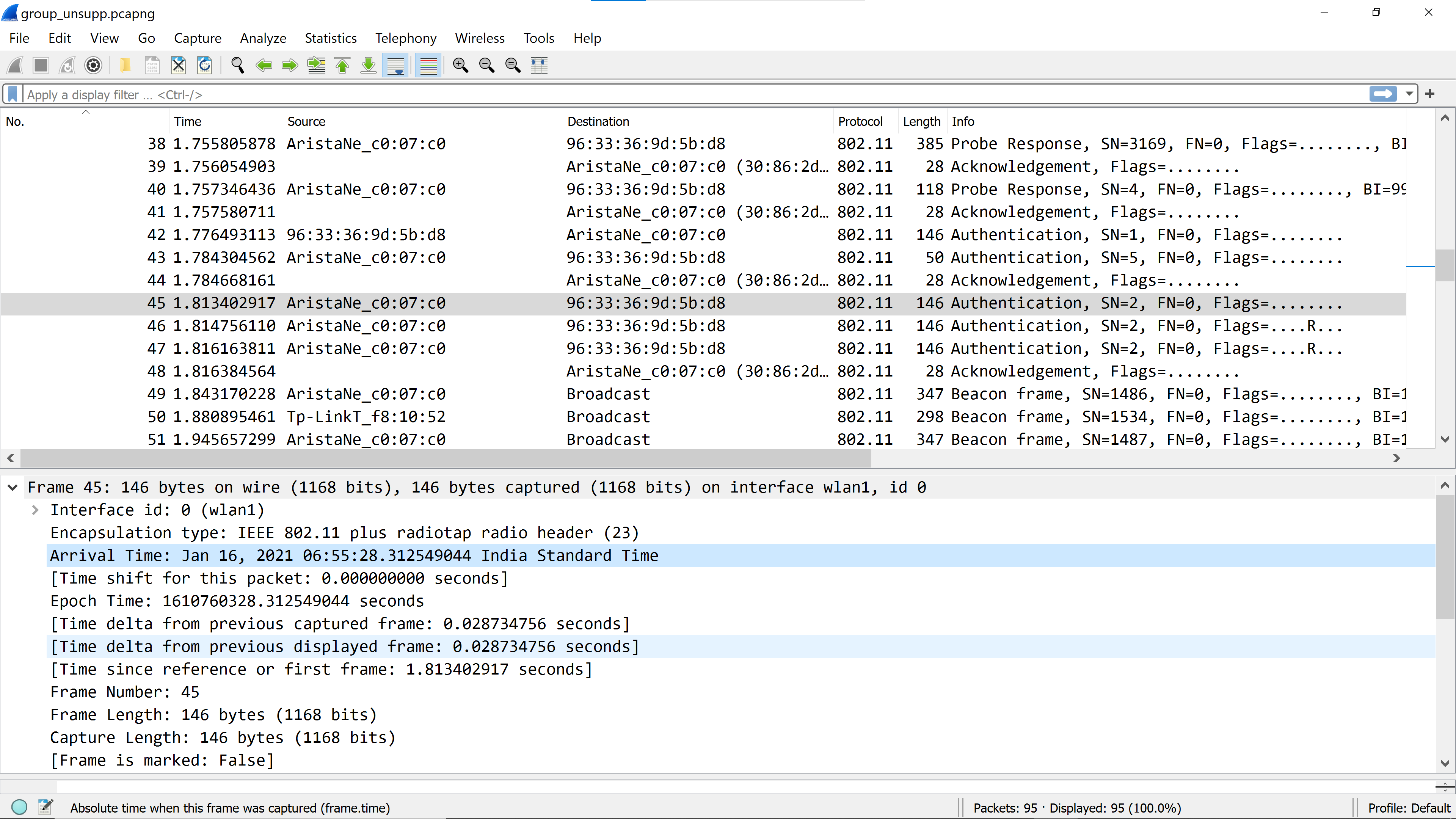}}
         %\caption{$W_{e}=W_{p}=W_{n}=1$}
%    	\label{Avg_Tp_Same_Wt} 
 \end{minipage}%
 ~
\begin{minipage}[b]{0.5\linewidth}
  \centering

  \resizebox{1.0\columnwidth}{!}{\includegraphics{./Figures/group_ids.png}}
    %\caption{$W_{e}=3$,$W_{p}=2$,$W_{n}=1$}
%    \label{Avg_Tp_Diff_Wt} 
\end{minipage}
    \caption{(a) Packet capture of SAE unsupported group DOS attack. Packet 43 is highlighted showing its status code as 0x004d (b) Same packet capture with packet 45 highlighted, showing its status code as successful (c) Same packet capture with packet 45 highlighted, showing its absolute arrival time (d) The response of IDS when this packet capture is given as input}
 \label{unsuppgroup}
%    \label{AT}
\end{figure}
}

\section{Conclusions and Future Work}\label{Con}
We launched and tested all the attacks on WPA3 networks that have been reported in prior work and two additional attacks using a testbed that contains an enterprise AP and an IDS. Our experimental results show that the AP is vulnerable to eight out of the above nine attacks  and the IDS is unable to detect any of them. We proposed a design for a signature-based IDS, which incorporates techniques to detect all the above attacks. Also, we implemented these techniques on our testbed and verified that our IDS is able to successfully detect all the above attacks. We provided schemes for mitigating the impact of the above attacks once they are detected.
We have made the code to perform the above attacks as well as that
of our IDS publicly available, so that it can be used for future
work by the research community at large. In particular, we believe that it will aid research in building a new and updated dataset based on WPA3, which in turn will help in developing improved IDSs.

\bibliography{BibFiles} 

\begin{thebibliography}{10}

\bibitem{Pahlavan2021}
K.~Pahlavan and P.~Krishnamurthy, ``{Evolution and Impact of Wi-Fi Technology
  and Applications: A Historical Perspective},'' {\em International Journal of
  Wireless Information Networks}, vol.~28, pp.~3--19, Mar 2021.

\bibitem{1997ieee}
``{IEEE Standard for Wireless LAN Medium Access Control (MAC) and Physical
  Layer (PHY) specifications},'' {\em IEEE Std 802.11-1997}, pp.~1--445, 1997.

\bibitem{wpa3}
``{Wi-Fi Alliance - WPA3 Specification Version 3.0, 2020}.''
  \url{https://www.wi-fi.org/discover-wi-fi/security }.
\newblock Accessed: 2021-06-06.

\bibitem{krack}
M.~Vanhoef and F.~Piessens, ``Key reinstallation attacks: Forcing nonce reuse
  in {WPA2},'' in {\em Proceedings of the 24th ACM Conference on Computer and
  Communications Security (CCS)}, ACM, 2017.

\bibitem{dos}
M.~Vanhoef and F.~Piessens, ``{Denial-of-Service Attacks Against the 4-way
  Wi-Fi Handshake},'' Ne{T}co{M}, 2017.

\bibitem{empirical}
C.~Kolias, G.~Kambourakis, A.~Stavrou, and S.~Gritzalis, ``{Intrusion Detection
  in 802.11 Networks: Empirical Evaluation of Threats and a Public Dataset},''
  {\em IEEE Communications Surveys Tutorials}, vol.~18, no.~1, pp.~184--208,
  2016.

\bibitem{dragon}
M.~Vanhoef and E.~Ronen, ``Dragonblood: Analyzing the {Dragonfly} handshake of
  {WPA3} and {EAP-pwd},'' in {\em IEEE Symposium on Security \& Privacy (SP)},
  IEEE, 2020.

\bibitem{depriv}
K.~Lounis and M.~Zulkernine, ``Wpa3 connection deprivation attacks,'' {\em In:
  Kallel S., Cuppens F., Cuppens-Boulahia N., Hadj Kacem A. (eds) Risks and
  Security of Internet and Systems. CRiSIS 2019. Lecture Notes in Computer
  Science}, vol.~12026, p.~164–176, 2020.

\bibitem{badtoken}
K.~Lounis and M.~Zulkernine, ``{Bad-Token: Denial of Service Attacks on
  WPA3},'' in {\em Proceedings of the 12th International Conference on Security
  of Information and Networks}, SIN '19, (New York, NY, USA), Association for
  Computing Machinery, 2019.

\bibitem{ids_need}
P.~Jadhav and S.~Boob, ``{Wireless Intrusion Detection System},'' {\em
  International Journal of Computer Applications}, vol.~5, no.~8, pp.~9--13,
  2010.

\bibitem{awid}
V.~L.~L. Thing, ``{IEEE} 802.11 {N}etwork anomaly detection and attack
  classification: {A} deep learning approach,'' in {\em {2017 IEEE Wireless
  Communications and Networking Conference (WCNC)}}, pp.~1--6, 2017.

\bibitem{awid2}
M.~E. Aminanto, R.~Choi, H.~C. Tanuwidjaja, P.~D. Yoo, and K.~Kim, ``{Deep
  Abstraction and Weighted Feature Selection for Wi-Fi Impersonation
  Detection},'' {\em {IEEE Transactions on Information Forensics and
  Security}}, vol.~13, no.~3, pp.~621--636, 2018.

\bibitem{semi}
J.~Ran, Y.~Ji, and B.~Tang, ``A semi-supervised learning approach to ieee
  802.11 network anomaly detection,'' in {\em 2019 IEEE 89th Vehicular
  Technology Conference (VTC2019-Spring)}, pp.~1--5, 2019.

\bibitem{nsl1}
N.~Shone, T.~N. Ngoc, V.~D. Phai, and Q.~Shi, ``A deep learning approach to
  network intrusion detection,'' {\em {IEEE Transactions on Emerging Topics in
  Computational Intelligence}}, vol.~2, no.~1, pp.~41--50, 2018.

\bibitem{nsl-kdd}
``{NSL-KDD} dataset.'' \url{https://www.unb.ca/cic/datasets/nsl.html}.
\newblock Accessed: 2021-06-06.

\bibitem{gprs1}
R.~Primartha and B.~A. Tama, ``Anomaly detection using random forest: A
  performance revisited,'' in {\em 2017 International Conference on Data and
  Software Engineering (ICoDSE)}, pp.~1--6, 2017.

\bibitem{aircrack}
``Aircrack-ng tool.'' \url{https://www.aircrack-ng.org/}.
\newblock Accessed: 2021-06-06.

\bibitem{mdk3}
``{KALI T}ools, {MDK3} tool.''
  \url{https://tools.kali.org/wireless-attacks/mdk3}.
\newblock Accessed: 2021-06-06.

\bibitem{metasploit}
``Metasploit penetration testing software.'' \url{https://www.metasploit.com/}.
\newblock Accessed: 2021-06-06.

\bibitem{advcomm}
M.~Vanhoef and F.~Piessens, ``{Advanced Wi-Fi Attacks Using Commodity
  Hardware},'' in {\em Proceedings of the 30th Annual Computer Security
  Applications Conference}, ACSAC '14, (New York, NY, USA), p.~256–265,
  Association for Computing Machinery, 2014.

\bibitem{gitlink}
N.~Dalal, ``{WPA3-Attacks-IDS}.''
  \url{https://github.com/neildalal/WPA3-Attacks-IDS}, 2021.

\bibitem{axel}
S.~Axelsson, ``{Intrusion detection systems: A survey and taxonomy},'' {\em
  Technical Report 99-15, Department of Computer Engineering, Chalmers
  University}, 2000.

\bibitem{wired2}
H.-J. Liao, C.-H. {Richard Lin}, Y.-C. Lin, and K.-Y. Tung, ``Intrusion
  detection system: A comprehensive review,'' {\em Journal of Network and
  Computer Applications}, vol.~36, no.~1, pp.~16--24, 2013.

\bibitem{ids1}
S.~Axelsson, ``{Research in Intrusion-Detection Systems: A Survey},'' The
  Swedish National Board for Industrial and Technical Development (NUTEK),
  1999.

\bibitem{ids2}
H.-J. Liao, C.-H. {Richard Lin}, Y.-C. Lin, and K.-Y. Tung, ``Intrusion
  detection system: A comprehensive review,'' {\em Journal of Network and
  Computer Applications}, vol.~36, no.~1, pp.~16--24, 2013.

\bibitem{history}
R.~Kemmerer and G.~Vigna, ``Intrusion detection: a brief history and
  overview,'' {\em Computer}, vol.~35, no.~4, pp.~supl27--supl30, 2002.

\bibitem{kdd}
``{KDD-CUP 1999} dataset.'' \url{http://kdd.ics.uci.edu/ databases}.
\newblock Accessed: 2021-06-06.

\bibitem{awid_21}
E.~Chatzoglou, G.~Kambourakis, and C.~Kolias, ``{Empirical Evaluation of
  Attacks Against IEEE 802.11 Enterprise Networks: The AWID3 Dataset},'' {\em
  IEEE Access}, vol.~9, pp.~34188--34205, 2021.

\bibitem{ids_tax}
H.~Debar, M.~Dacier, and A.~Wespi, ``Towards a taxonomy of intrusion-detection
  systems,'' {\em Computer Networks}, vol.~31, no.~8, pp.~805--822, 1999.

\bibitem{snort}
``Snort-{W}ireless.'' \url{https://www.snort.org/}.
\newblock Accessed: 2021-06-06.

\bibitem{airmag}
``{Air{M}agnet (Now called NetAlly)}.'' \url{http://www.airmagnet.com}.
\newblock Accessed: 2021-06-06.

\bibitem{airdef}
``Extreme {A}ir{D}efence.'' \url{https://www.extremenetworks.com/}.
\newblock Accessed: 2021-06-06.

\bibitem{802112016}
``{IEEE} standard for information technology--telecommunications and
  information exchange between systems local and metropolitan area
  networks--specific requirements {P}art 11: {W}ireless {LAN} medium access
  control ({MAC}) and physical layer ({PHY}) specifications,'' in {\em IEEE Std
  802.11-2012 (Revision of IEEE Std 802.11-2007}, p.~1–2793.

\bibitem{wep1}
N.~Cam-Winget, R.~Housley, D.~Wagner, and J.~Walker, ``{Security Flaws in
  802.11 Data Link Protocols},'' {\em Communications of the ACM}, vol.~46,
  no.~5, p.~35–39.

\bibitem{wep2}
A.~Bittau, M.~Handley, and J.~Lackey, ``{The final nail in WEP's coffin},'' in
  {\em 2006 IEEE Symposium on Security and Privacy (S P'06)}, pp.~15 pp.--400,
  2006.

\bibitem{wep3}
S.~Fluhrer, I.~Mantin, and A.~Shamir, ``Weaknesses in the key scheduling
  algorithm of rc4,'' in {\em Selected Areas in Cryptography} (S.~Vaudenay and
  A.~M. Youssef, eds.), (Berlin, Heidelberg), pp.~1--24, Springer Berlin
  Heidelberg, 2001.

\bibitem{wep4}
P.~Sepehrdad, S.~Vaudenay, and M.~Vuagnoux, ``Discovery and exploitation of new
  biases in rc4,'' in {\em Selected Areas in Cryptography} (A.~Biryukov,
  G.~Gong, and D.~R. Stinson, eds.), (Berlin, Heidelberg), pp.~74--91, Springer
  Berlin Heidelberg, 2011.

\bibitem{80211w}
``{IEEE Standard for Information technology - Telecommunications and
  information exchange between systems - Local and metropolitan area networks -
  Specific requirements. Part 11: Wireless LAN Medium Access Control (MAC) and
  Physical Layer (PHY) Specifications Amendment 4: Protected Management
  Frames},'' in {\em IEEE Std 802.11w-2009}, pp.~1–111,.

\bibitem{deauthattack}
J.~Bellardo and S.~Savage, ``{802.11 Denial-of-Service Attacks: Real
  Vulnerabilities and Practical Solutions},'' in {\em Proceedings of the 12th
  Conference on USENIX Security Symposium - Volume 12}, SSYM'03, (USA), p.~2,
  USENIX Association, 2003.

\bibitem{wpa2attack}
O.~Nakhila, A.~Attiah, Y.~Jin, and C.~Zou, ``Parallel active dictionary attack
  on wpa2-psk wi-fi networks,'' in {\em MILCOM 2015 - 2015 IEEE Military
  Communications Conference}, pp.~665--670, 2015.

\bibitem{vink}
M.~Vink, E.~Poll, and A.~Verbiest, ``{A Comprehensive Taxonomy of Wi-Fi
  Attacks},'' in {\em Master Thesis Cyber Security, Radboud University
  Nijmegen}, 2020.

\bibitem{dragondrain}
``{GitHub - Dragondrain and Dragontime by Vanhoefm}.''
  \url{https://github.com/vanhoefm/dragondrain-and-time }.
\newblock Accessed: 2021-06-06.

\bibitem{hostapd}
``Hostapd.'' \url{https://w1.fi/hostapd/}.
\newblock Accessed: 2021-06-06.

\bibitem{wpasupp}
``wpa\_supplicant.'' \url{https://w1.fi/wpa_supplicant/}.
\newblock Accessed: 2021-06-06.

\bibitem{nist}
P.~Karen~Scarfone, Mell, ``Guide to intrusion detection and prevention systems
  (idps),'' {\em Computer Security Resource Center, Recommendations of the
  National Institute of Standards and Technology}, Feb 2007.

\end{thebibliography}
\bibliographystyle{ieeetr}

\end{document}